\documentclass[twocolumn]{aastex62}
\usepackage{captcont,xcolor}
\usepackage{verbatim}
\usepackage{hyperref}
\usepackage{multirow}

\usepackage{CJKutf8}

\shorttitle{ALMA-Taurus}
\shortauthors{Long et al.}

\begin{document}
\begin{CJK*}{UTF8}{gbsn}

\title{Compact Disks in a High Resolution ALMA Survey of Dust Structures in the Taurus Molecular Cloud}


\author{Feng Long(龙凤)}
\affiliation{Kavli Institute for Astronomy and Astrophysics, Peking University, Beijing 100871, China}
\affiliation{Department of Astronomy, School of Physics, Peking University,  Beijing 100871, China}

\author{Gregory J. Herczeg(沈雷歌)}
\affiliation{Kavli Institute for Astronomy and Astrophysics, Peking University, Beijing 100871, China}

\author{Daniel Harsono}
\affiliation{Leiden Observatory, Leiden University, P.O. box 9513, 2300 RA Leiden, The Netherlands}

\author{Paola Pinilla}
\affiliation{Department of Astronomy/Steward Observatory, The University of Arizona, 933 North Cherry Avenue, Tucson, AZ 85721, USA}
\affiliation{Max-Planck-Institut f\"{u}r Astronomie, K\"{o}nigstuhl 17, 69117, Heidelberg, Germany}

\author{Marco Tazzari}
\affiliation{Institute of Astronomy, University of Cambridge, Madingley Road, Cambridge CB3 0HA, UK}

\author{Carlo F. Manara}
\affiliation{European Southern Observatory, Karl-Schwarzschild-Str. 2, D-85748 Garching bei M\"{u}nchen, Germany}

\author{Ilaria Pascucci}
\affiliation{Lunar and Planetary Laboratory, University of Arizona, Tucson, AZ 85721, USA}
\affiliation{Earths in Other Solar Systems Team, NASA Nexus for Exoplanet System Science, USA}

\author{Sylvie Cabrit}
\affiliation{Sorbonne Universit\'{e}, Observatoire de Paris, Universit\'{e} PSL, CNRS, LERMA, F-75014 Paris, France}
\affiliation{Univ. Grenoble Alpes, CNRS, IPAG, F-38000 Grenoble, France}

\author{Brunella Nisini}
\affiliation{INAF–Osservatorio Astronomico di Roma, via di Frascati 33, 00040 Monte Porzio Catone, Italy}

\author{Doug Johnstone}
\affiliation{NRC Herzberg Astronomy and Astrophysics, 5071 West Saanich Road, Victoria, BC, V9E 2E7, Canada}
\affiliation{Department of Physics and Astronomy, University of Victoria, Victoria, BC, V8P 5C2, Canada}

\author{Suzan Edwards}
\affiliation{Five College Astronomy Department, Smith College, Northampton, MA 01063, USA}

\author{Colette Salyk}
\affiliation{Vassar College Physics and Astronomy Department, 124 Raymond Avenue, Poughkeepsie, NY 12604, USA}

\author{Francois Menard}
\affiliation{Univ. Grenoble Alpes, CNRS, IPAG, F-38000 Grenoble, France}

\author{Giuseppe Lodato}
\affiliation{Dipartimento di Fisica, Universita Degli Studi di Milano, Via Celoria, 16, I-20133 Milano, Italy}

\author{Yann Boehler}
\affiliation{Univ. Grenoble Alpes, CNRS, IPAG, F-38000 Grenoble, France}
\affiliation{Rice University, Department of Physics and Astronomy, Main Street, 77005 Houston, USA}

\author{Gregory N. Mace}
\affiliation{McDonald Observatory and Department of Astronomy, University of Texas at Austin, 2515 Speedway, Stop C1400, Austin, TX 78712-1205, USA}

\author{Yao Liu}
\affiliation{Max-Planck-Institut f\"{u}r Extraterrestrische Physik, Giessenbachstrasse 1, 85748, Garching, Germany}
\affiliation{Purple Mountain Observatory, Chinese Academy of Sciences, 2 West Beijing Road, Nanjing 210008, China}

\author{Gijs D. Mulders}
\affiliation{Department of the Geophysical Sciences, The University of Chicago, 5734 South Ellis Avenue, Chicago, IL 60637, USA}
\affiliation{Earths in Other Solar Systems Team, NASA Nexus for Exoplanet System Science, USA}

\author{Nathanial Hendler}
\affiliation{Lunar and Planetary Laboratory, University of Arizona, Tucson, AZ 85721, USA}
\affiliation{LSSTC Data Science Fellow}

\author{Enrico Ragusa}
\affiliation{Department of Physics and Astronomy, University of Leicester, Leicester LE1 7RH, UK}

\author{William J. Fischer}
\affiliation{Space Telescope Science Institute Baltimore, MD 21218, USA}

\author{Andrea Banzatti}
\affiliation{Lunar and Planetary Laboratory, University of Arizona, Tucson, AZ 85721, USA}

\author{Elisabetta Rigliaco}
\affiliation{INAF-Osservatorio Astronomico di Padova, Vicolo dell'Osservatorio 5, 35122 Padova, Italy}

\author{Gerrit van de Plas}
\affiliation{Univ. Grenoble Alpes, CNRS, IPAG, F-38000 Grenoble, France}

\author{Giovanni Dipierro}
\affiliation{Department of Physics and Astronomy, University of Leicester, Leicester LE1 7RH, UK}

\author{Michael Gully-Santiago}
\affiliation{NASA Ames Research Center and Bay Area Environmental Research Institute, Moffett Field, CA 94035, USA}

\author{Ricardo Lopez-Valdivia}
\affiliation{McDonald Observatory and Department of Astronomy, University of Texas at Austin, 2515 Speedway, Stop C1400, Austin, TX 78712-1205, USA}

\begin{abstract}
We present a high-resolution ($\sim0\farcs12$, $\sim16$ au, mean sensitivity of $50~\mu$Jy~beam$^{-1}$ at 225 GHz) snapshot survey of 32 protoplanetary disks around young stars with spectral type earlier than M3 in the Taurus star-forming region using Atacama Large Millimeter Array (ALMA). This sample includes most mid-infrared excess members that were not previously imaged at high spatial resolution, excluding close binaries and highly extincted objects, thereby providing a more representative look at disk properties at 1--2 Myr. Our 1.3 mm continuum maps reveal 12 disks with prominent dust gaps and rings, 2 of which are around primary stars in wide binaries, and 20 disks with no resolved features at the observed resolution (hereafter smooth disks), 8 of which are around the primary star in wide binaries. The smooth disks were classified based on their lack of resolved substructures, but their most prominent property is that they are all compact with small effective emission radii ($R_{\rm eff,95\%} \lesssim 50$ au).  In contrast, all disks with $R_{\rm eff,95\%}$ of at least 55 au in our sample show detectable substructures. Nevertheless, their inner emission cores (inside the resolved gaps) have similar peak brightness, power law profiles, and transition radii to the compact smooth disks, so the primary difference between these two categories is the lack of outer substructures in the latter. These compact disks may lose their outer disk through fast radial drift without dust trapping, or they might be born with small sizes. The compact dust disks, as well as the inner disk cores of extended ring disks, that look smooth at the current resolution will likely show small-scale or low-contrast substructures at higher resolution. The correlation between disk size and disk luminosity correlation demonstrates that some of the compact disks are optically thick at millimeter wavelengths. 
\end{abstract}

\section{Introduction}
The rich diversity in exoplanetary systems \citep[see review by][]{winn2015} must have its origin, at least in part, when planets are still forming in their natal protoplanetary disks.  It is therefore not surprising that protoplanetary disks also show spectacular diversity in virtually every observable disk property. This diversity was initially seen in the decades-old problem of why some disks survive for $>10$ Myr while others disappear in $<1$ Myr \citep[e.g.][]{walter1988,skrutskie90,haisch2001}. In recent ALMA surveys, disks in each stellar mass bin have a spread in disk dust mass of $\sim 2$ orders of magnitude \citep{barenfeld2016,pascucci2016,ansdell2016,ansdell2017}.  Similar spreads are seen in stellar accretion rates \citep[e.g.,][]{manara2017} and in disk CO gas masses \citep{miotello2017,long2017}.  This diversity is also now being seen at high-spatial resolution, with remarkable images that reveal an assortment of rings, cavities, spirals, and horseshoe-like substructures in both millimeter continuum emission (e.g., \citealt{alma2015, andrews2016, Perez2016}) and near-IR scattered light observations from small dust grains \citep[e.g.][]{vanBoekel2017,avenhaus2018,garufi2018}.

An emerging view is that substructures of mm-sized grains are identified in most disks, when they are imaged  with sufficient angular resolution \citep{vanderMarel2013, isella2016, cieza2017, loomis2017, vanderplas2017,Hendler2018, fedele18,boehler2018, dong2018, vanTerwisga2018,vandermarel2019}. These substructures may be either a cause or a consequence of planetesimal and planet formation. However, the frequency of such structures has been uncertain because deep, high-spatial resolution ALMA observations so far have preferentially targeted stars with known large dust cavities and the brightest known disks. These biases developed naturally because transition disks (disks with inner cavities) are a likely signature of planet formation, while the brightest disks are easier to observe at the highest spatial resolutions.

Several recent programs have sought to minimize selection biases by obtaining high-resolution imaging of more complete samples.  Deep imaging of 20 of the brightest disks in Lupus, Ophiuchus, and Upper Sco at $\sim0\farcs03$ resolution revealed that rings are very common, while spiral arms and other asymmetric structures are rare \citep[e.g.][]{andrews2018_dsharp,huang2018_ring,huang2018_sprial}. Meanwhile, in the first results of 147 disks in a much broader survey of Ophiucus with $\sim 0\farcs2$ resolution, \citet{cieza19} finds that most disks are small ($<15$ AU), in contrast to the picture of large rings that has emerged from brightness-selected samples.

In this paper, we present the overview of the properties of dust disks in high-resolution ($\sim 0\farcs12$) ALMA imaging of 32 protoplanetary disks in the Taurus Molecular Cloud, selected to be representative of disks across a wide range of sub-mm flux and not selected for previous identification of inner holes from near- and mid-IR spectral energy distributions. This survey was designed with sufficient resolution and depth to provide a snapshot of substructures of mm-sized grains in a large number of disks.  In initial results from our survey, we described the detected substructures in our sample and used them to rule out the  hypothesis that they are all generated by ice lines \citep{long2018}, evaluated and modeled the prominent ring around MWC 480 \citep{liu2019}, and identified the gap-inferred young planet population, under the assumption that the gaps are carved by planets \citep{lodato2019}. A companion paper by Manara et al.~(submitted) further evaluates the disks in resolved binary systems in our sample.  Here we present an analysis of the full sample, with an emphasis on those disks around single stars that did not have resolved substructures identified by \citet{long2018}.   In Section~\ref{sec:sample&obs}, we describe the sample, including how the targets were selected, and the ALMA observations. In Section~\ref{sec:model}, we characterize disk properties by fitting the observations in the  visibility plane. In Section~\ref{sec:result}, we examine the commonalities and differences in stellar and disk properties for disks with different dust morphologies. In Section~\ref{sec:discussion}, we discuss the future directions towards detecting disk substructures.  We close with our main findings in Section~\ref{sec:summary}.

\section{Sample and Observations} \label{sec:sample&obs}

\subsection{Sample Selection} \label{sec:sample}

The goal of our target selection was to obtain a sample that spans the full range of disk types for solar-mass stars, without any bias related to any disk property.  Previous measurements of the disks, including disk brightness and inference of substructures from SEDs, were explicitly not used in the target selection, except for a previous identification of a primordial disk.

Our sample selection began with the census of Taurus disks around stars identified by Spitzer \citep{Rebull10,luhman2010}.  We selected disks around stars of spectral type earlier than M3 to ensure sufficient signal-to-noise to image disks across the full range of disk brightness at our sensitivity.  Known binaries with separations between $0\farcs1$--$0\farcs5$ were excluded to avoid interactions at our spatial resolution. 
Sources with high extinction ($A_V>3$ mag) or consistently faint optical/near-IR emission were excluded to avoid edge-on disks and embedded objects.  We also excluded from our sample all disks with existing (or scheduled) ALMA images of dust emission with a spatial resolution better than $0\farcs25$.
This avoidance of near-duplications is the most significant bias that introduces uncertainties in making robust generalizations from our current sample.  Many of the most well-known disks had existing high-resolution observations at the time of our proposal.  The final selection eliminated two isolated targets to optimize the efficiency of the ALMA observing blocks.  A more complete description of targets that were excluded from our sample is described in Appendix~\ref{sec:source-selection}.

\begin{figure}[!t]
\centering
     \includegraphics[width=0.49\textwidth]{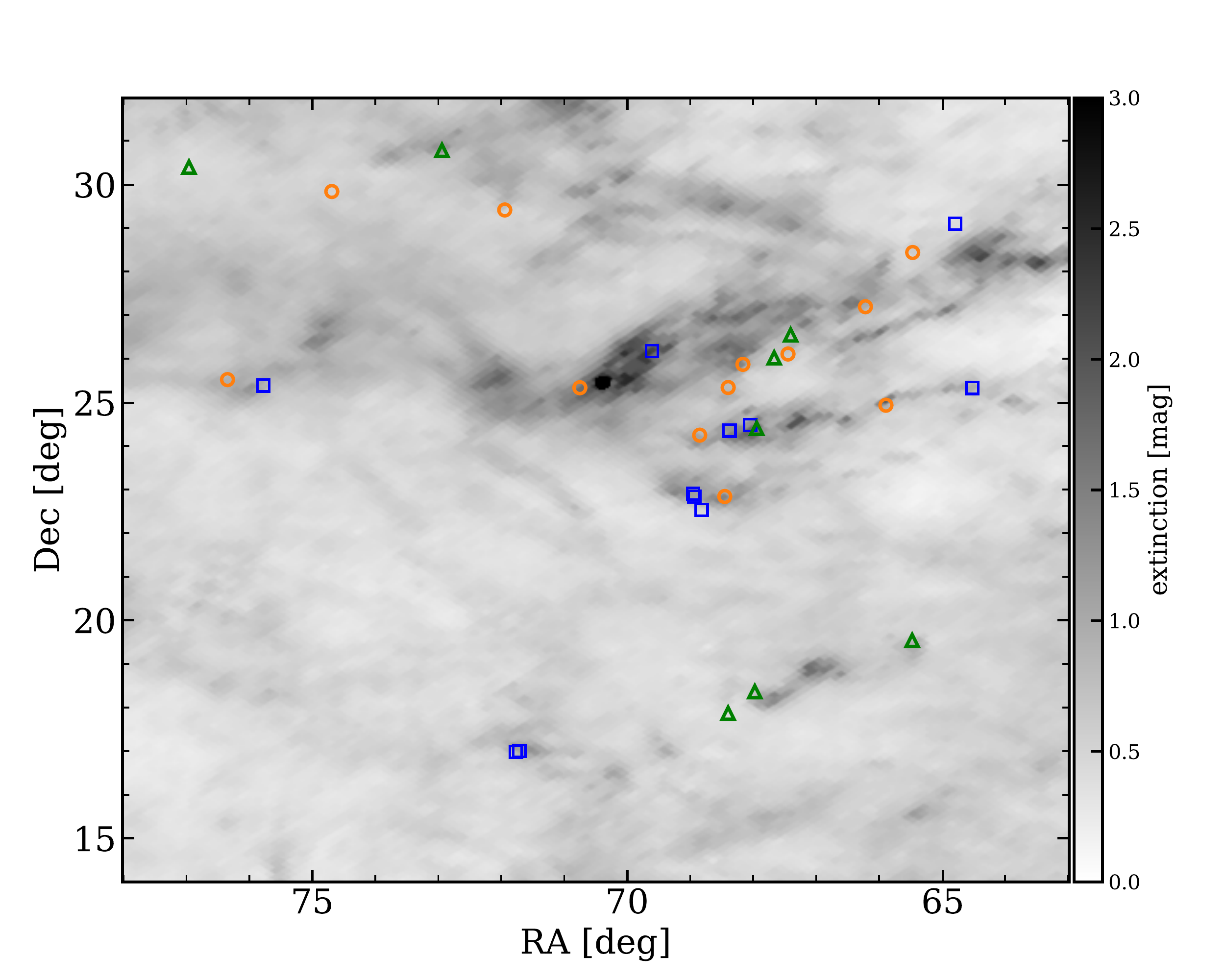} \\
    \caption{The spatial distribution of the 32 disks in Taurus Clouds selected for our ALMA Survey. Disks with substructures are shown in orange, while smooth disks in singles and in binaries are shown in blue and green, respectively (see the sub-sample category in \S~\ref{sec: category}). The background is an extinction map compiled by \citet{schlafly2014}, in which some missing data in the densest region are filled with A$_V$=2. \label{fig:spatial} }
\end{figure}

These selection criteria produced a sample of 32 stellar systems, including 10 systems in wide binaries.  The spatial distribution of these systems (Figure~\ref{fig:spatial}) shows that the sources are located across the Taurus Molecular Cloud, with the densest parts of the cloud excluded because of criterion that required low extinction.

\begin{deluxetable*}{lcccccccccccc}
\tabletypesize{\scriptsize}
\tablecaption{Host Stellar Properties and Observation Results \label{tab:source_prop}}
\tablewidth{0pt}
\tablehead{
\colhead{Name} & \colhead{2MASS} & \colhead{D} &\colhead{A$_V$}\tablenotemark{a} & \colhead{SpTy} & \colhead{$T_{\rm eff}$} & \colhead{$L_*$} & \colhead{$M_*$} & \colhead{$t_*$} & \colhead{Multiplicity} & \colhead{Peak $I_{\nu}$} & \colhead{RMS noise} & \colhead{beam} \\
\colhead{} & \colhead{} & \colhead{(pc)} & \colhead{(mag)} & \colhead{} & \colhead{(K)} & \colhead{($L_\odot$)} &  \colhead{($M_\odot$)} & \colhead{(Myr)} & \colhead{(arcsec)} & \colhead{(mJy beam$^{-1}$)} & \colhead{($\mu$Jy beam$^{-1}$)} & \colhead{(arcsec)} \\
} 
\colnumbers
\startdata
\multicolumn{13}{c}{disks with substructures} \\ 
\hline
    CI Tau & 04335200+2250301 &  158 &   1.90 & K5.5 & 4277 &  0.81 & 0.89$^{+0.21}_{-0.17}$  &  2.50$^{+ 2.00}_{-1.10}$ &         -- &  8.55 &  50 & 0.13$\times$0.11 \\
  CIDA 9 A & 05052286+2531312 &  171 &  1.35 & M1.8 & 3589 &  0.20 & 0.43$^{+0.15}_{-0.10}$  &  3.20$^{+ 3.10}_{-1.60}$ &  2.34(K11) &  2.98 &  50 & 0.13$\times$0.10 \\
    DL Tau & 04333906+2520382 &  159 &   1.80 & K5.5 & 4277 &  0.65 & 0.98$^{+0.84}_{-0.15}$  &  3.50$^{+ 2.80}_{-1.60}$ &         -- & 12.27 &  49 & 0.14$\times$0.11 \\
    DN Tau & 04352737+2414589 &  128 &  0.55 & M0.3 & 3806 &  0.70 & 0.52$^{+0.14}_{-0.11}$  &  0.90$^{+ 0.60}_{-0.40}$ &         -- & 12.87 &  51 & 0.13$\times$0.11 \\
    DS Tau & 04474859+2925112 &  159 &  0.25 & M0.4 & 3792 &  0.25 & 0.58$^{+0.17}_{-0.13}$  &  4.80$^{+ 4.80}_{-2.30}$ &         -- &  3.05 &  50 & 0.14$\times$0.10 \\
    FT Tau & 04233919+2456141 &  127 &   1.30 & M2.8 & 3444 &  0.15 & 0.34$^{+0.13}_{-0.09}$  &  3.20$^{+ 3.20}_{-1.60}$ &         -- & 10.76 &  48 & 0.12$\times$0.11 \\
    GO Tau & 04430309+2520187 &  144 &   1.50 & M2.3 & 3516 &  0.21 & 0.36$^{+0.13}_{-0.09}$  &  2.20$^{+ 1.90}_{-1.10}$ &         -- &  7.87 &  49 & 0.14$\times$0.11 \\
    IP Tau & 04245708+2711565 &  130 &  0.75 & M0.6 & 3763 &  0.34 & 0.52$^{+0.15}_{-0.13}$  &  2.50$^{+ 2.20}_{-1.20}$ &         -- &  1.66 &  48 & 0.14$\times$0.11 \\
    IQ Tau & 04295156+2606448 &  131 &  0.85 & M1.1 & 3690 &  0.22 & 0.50$^{+0.16}_{-0.12}$  &  4.20$^{+ 4.10}_{-2.00}$ &         -- &  5.76 &  78 & 0.16$\times$0.11 \\
   MWC 480 & 04584626+2950370 &  161 &   0.10 & A4.5 & 8400 & 17.38 & 1.91$^{+0.09}_{-0.13}$  &  6.90$^{+ 5.10}_{-5.80}$ &         -- & 31.29 &  69 & 0.17$\times$0.11 \\
    RY Tau & 04215740+2826355 &  128 &  1.95 &   F7 & 6220 & 12.30 & 2.04$^{+0.30}_{-0.26}$  &  5.00$^{+ 3.10}_{-1.60}$ &         -- & 18.98 &  51 & 0.14$\times$0.11 \\
  UZ Tau E\tablenotemark{b} & 04324303+2552311 &  131 &   0.90 & M1.9 & 3574 &  0.35 & 1.23$\pm$0.07   &  (1.30$^{+ 1.00}_{-0.60}$) &  3.54(K09) &  8.44 &  49 & 0.13$\times$ 0.1 \\
\hline
\multicolumn{13}{c}{smooth disks in single stars} \\ 
\hline
    BP Tau & 04191583+2906269 &  129 &  0.45 & M0.5 & 3777 &  0.40 & 0.52$^{+0.15}_{-0.12}$  &  1.90$^{+ 1.50}_{-0.90}$ &         -- &  5.18 &  45 & 0.14$\times$0.11 \\
    DO Tau & 04382858+2610494 &  139 &  0.75 & M0.3 & 3806 &  0.23 & 0.59$^{+0.15}_{-0.13}$  &  5.90$^{+ 6.10}_{-2.80}$ &         -- & 22.67 &  58 & 0.14$\times$0.10 \\
    DQ Tau\tablenotemark{c} & 04465305+1700001 &  197 &   1.40 & M0.6 & 3763 &  1.17 & 1.61$^{+0.58}_{-0.34}$  &  (0.5)                      &         -- & 23.05 &  45 & 0.13$\times$0.10 \\
    DR Tau & 04470620+1658428 &  195 &  0.45 &   K6 & 4205 &  0.63 & 0.93$^{+0.85}_{-0.16}$  &  3.20$^{+ 2.70}_{-1.40}$ &         -- & 21.11 &  51 & 0.13$\times$0.10 \\
    GI Tau & 04333405+2421170 &  130 &  2.05 & M0.4 & 3792 &  0.49 & 0.52$^{+0.15}_{-0.12}$  &  1.50$^{+ 1.20}_{-0.70}$ &         -- &  4.33 &  50 & 0.12$\times$0.11 \\
    GK Tau & 04333456+2421058 &  129 &   1.50 & K7.5 & 4007 &  0.80 & 0.63$^{+0.16}_{-0.13}$  &  1.20$^{+ 0.70}_{-0.60}$ &         -- &  3.26 &  51 & 0.12$\times$0.11 \\
 Haro 6-13 & 04321540+2428597 &  130 &  2.25 & K5.5 & 4277 &  0.79 & 0.91$^{+0.24}_{-0.17}$  &  2.60$^{+ 2.10}_{-1.10}$ &         -- & 32.63 &  52 & 0.14$\times$0.11 \\
    HO Tau & 04352020+2232146 &  161 &   1.00 & M3.2 & 3386 &  0.14 & 0.30$^{+0.05}_{-0.04}$  &  2.70$^{+ 1.50}_{-1.00}$ &         -- &  3.93 &  46 & 0.12$\times$0.11 \\
    HP Tau & 04355277+2254231 &  177 &  3.15 & K4.0 & 4590 &  1.30 & 1.20$^{+1.14}_{-0.18}$  &  2.40$^{+ 1.90}_{-1.00}$ &         -- & 22.45 &  51 & 0.13$\times$0.11 \\
    HQ Tau & 04354733+2250216 &  158 &   2.60 & K2.0 & 4900 &  4.34 & 1.78$^{+1.69}_{-0.26}$  &  1.00$^{+ 0.60}_{-0.40}$ &         -- &  1.16 &  46 & 0.12$\times$0.11 \\
  V409 Tau & 04181078+2519574 &  131 &   1.00 & M0.6 & 3763 &  0.66 & 0.50$^{+0.13}_{-0.10}$  &  0.90$^{+ 0.50}_{-0.30}$ &         -- &  4.48 &  46 & 0.13$\times$0.11 \\
  V836 Tau & 05030659+2523197 &  169 &   0.60 & M0.8 & 3734 &  0.44 & 0.48$^{+0.14}_{-0.12}$  &  1.40$^{+ 1.10}_{-0.70}$ &         -- &  7.64 &  49 & 0.13$\times$0.10 \\
\hline
\multicolumn{13}{c}{smooth disks around the primary star in binaries/multiple systems} \\
\hline
  DH Tau A & 04294155+2632582 &  135 &  0.65 & M2.3 & 3516 &  0.20 & 0.37$^{+0.13}_{-0.10}$  &  2.30$^{+ 2.10}_{-1.20}$ &  2.34(I05) &  9.14 &  44 & 0.13$\times$0.11 \\
  DK Tau A & 04304425+2601244 &  128 &   0.70 & K8.5 & 3902 &  0.45 & 0.60$^{+0.16}_{-0.13}$  &  2.30$^{+ 1.80}_{-1.10}$ &  2.36(KH09) & 12.73 &  44 & 0.13$\times$0.11 \\
  HK Tau A & 04315056+2424180 &  133 &   2.40 & M1.5 & 3632 &  0.27 & 0.44$^{+0.14}_{-0.11}$  &  2.20$^{+ 1.90}_{-1.10}$ &  2.34(KH09) & 11.56 &  48 & 0.12$\times$0.11 \\
  HN Tau A\tablenotemark{d} & 04333935+1751523 &  136 &  1.15 &   K3 & 4730 &  0.16 & 1.53$\pm$0.15           &  (2.0)                      &  3.14(KH09) &   7.0 &  40 & 0.14$\times$0.10 \\
  RW Aur A & 05074953+3024050 &  163 & (0) &   K0 & 5250 &  0.99 & 1.20$^{+0.18}_{-0.13}$  & 13.50$^{+11.10}_{-5.90}$ & 1.42(WG01) & 18.34 &  51 & 0.16$\times$0.10 \\
   T Tau N & 04215943+1932063 &  144 &  1.25 &   K0 & 5250 &  6.82 & 2.19$^{+0.38}_{-0.24}$  &  1.10$^{+ 0.70}_{-0.40}$ &  0.68(KH09) & 64.56 &  52 & 0.14$\times$0.10 \\
  UY Aur A & 04514737+3047134 &  155 &   1.00 & K7.0 & 4060 &  1.05 & 0.65$^{+0.17}_{-0.13}$  &  0.90$^{+ 0.40}_{-0.40}$ &  0.88(KH09) & 16.91 &  48 & 0.15$\times$0.10 \\
V710 Tau A\tablenotemark(e) & 04315779+1821350 &  142 &  0.55 & M1.7 & 3603 &  0.26 & 0.42$^{+0.13}_{-0.11}$  &  2.20$^{+ 1.90}_{-1.10}$ &  3.22(KH09) &  7.52 &  42 & 0.14$\times$0.10 \\
\enddata
\tablecomments{Our sample is divided into three sub-groups (as listed in the Table with three segments), from top to bottom. The distance for individual star is adopted from the Gaia DR2 parallax \citep{gaia2018}. 
Spectral type is adopted from \citet{herczeg2014} and stellar luminosity is calculated from J-band magnitude and updated to the new Gaia distance. Stellar mass and age are re-calculated with the stellar luminosity and effective temperature listed here using the same method as in \citet{pascucci2016}. The last three columns list the peak intensity in continuum maps, noise level, and synthesised beam FWHM.}
\tablenotetext{a}{$A_V$ is listed to nearest 0.05 and has an uncertainty of $\sim 0.2-0.5$ mag; the higher uncertainty applies to stars with high veiling at optical wavelengths.  RW Aur has a negative statistical extinction and is treated as $A_V=0$ mag here.}
\tablenotetext{b}{UZ Tau E is a spectroscopic binary in 0.03 au separation \citep{mathieu1996, prato2002}. We adopt its stellar mass from dynamical measurement \citep{simon00}.}
\tablenotetext{c}{DQ Tau is a double-lined spectroscopic binary with a period of $\sim$16 days in an ecentric orbit (e = 0.56, \citealt{mathieu1997, tofflemire2017}). Its stellar mass is adopted from the dynamical measurement of \citet{czekala2016}.} 
\tablenotetext{d}{HN Tau A has a high inclination angle and appears too faint to derive the accurate stellar mass and age from the grids, for which we adopt the dynamical mass measurement from \citet{simon2017}.}
\tablenotetext{e}{V710 Tau North, see discussion of nomenclature in Manara et al.~submitted.}
\tablerefs{The references for quoted stellar multiplicity: K11=\citet{kraus2011}, I05=\citet{itoh2005}, KH09=\citet{kraus2009}, WG01=\citet{white2001}. }

\end{deluxetable*}

\subsection{Host star properties} \label{sec:host-star}

Table~\ref{tab:source_prop} lists the properties of the host stars in our ALMA sample.  Most spectral types and the spectral type-temperature conversion are obtained from the optical spectral survey of \citet{herczeg2014}.  Luminosities are then calculated from the 2MASS $J$-band magnitude \citep{skrutskie2006}, the extinction measured by \citet{herczeg2014}, the $J$-band bolometric correction for the relevant spectral type calculated by \citet{pecaut13}, and the distance from {\it Gaia} DR2 \citep{gaia2018}.  The properties of RY Tau were unclear from literature estimates and are derived in Appendix~\ref{sec:source-detail}.

\begin{figure}[!t]
\centering
    \includegraphics[width=0.49\textwidth]{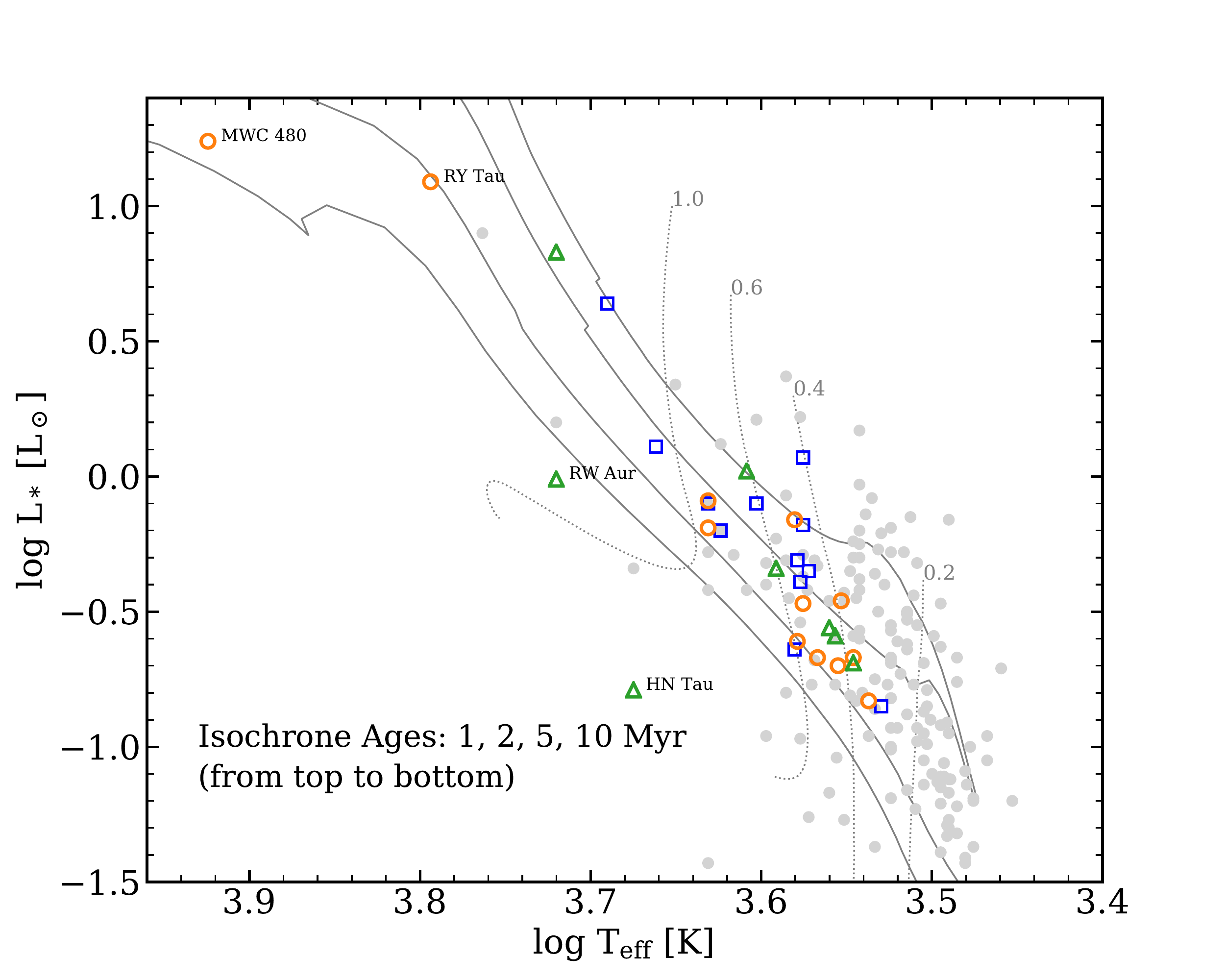} \\
    \caption{HR diagram of Taurus sources. Our ALMA sample is labeled with colors as Figure~\ref{fig:spatial}, while the other Taurus members listed in \citet{andrews2013} are shown in grey dots. We use the non-magnetic evolutionary tracks from \citet{feiden2016} to cover our ALMA sample, with grey dotted lines representing evolutionary tracks for different stellar masses.   \label{fig:HRD}}
\end{figure}

\begin{deluxetable*}{lccccc}
\tabletypesize{\scriptsize}
\tablecaption{ALMA Observing Log\label{tab:ALMAObservations}}
\tablehead{
\colhead{UTC Date} & 
\colhead{N$_{\rm ant}$} & 
\colhead{Baselines/m} & 
\colhead{PWV/mm} & 
\colhead{Calibrators} & 
\colhead{Targets} 
}
\colnumbers
\startdata
2017/08/18  &   43  & 21-3638   & 0.5 & J0423-0120,J0423-0120,J0431+1731 & T Tau, HN Tau, V710 Tau \\
            &       &           &     & J0423-0120,J0423-0120,J0440+1437 & DQ Tau, DR Tau \\
\hline
2017/08/27  &   47  & 21-3638   & 0.5 & J0510+1800,J0510+1800,J0512+2927 & MWC 480 \\
            &       &           &     & J0510+1800,J0510+1800,J0435+2532\tablenotemark{*} &  CI Tau, DL Tau, DN Tau, HP Tau, Haro 6-13, RY Tau \\
            &       &           &     & J0510+1800,J0510+1800,J0440+2728 & DO Tau, GO Tau \\
            &       &           &     & J0510+1800,J0510+1800,J0426+2327 & IQ Tau \\
\hline
2017/08/31  &   45  & 21-3697   & 1.3 & J0510+1800,J0510+1800,J0519+2744 & V836 Tau, CIDA 9\\
            &       &           &     & J0510+1800,J0510+1800,J0439+3045 & UY Aur, DS Tau \\
            &       &           &     & J0510+1800,J0510+1800,J0512+2927 & RW Aur \\
\hline
2017/08/31  &   45  & 21-3697   & 1.5 & J0510+1800,J0423-0120,J0426+2327 & DK Tau, GK Tau, V409 Tau, GI Tau,  FT Tau\\
			&       &           &     &  & HO Tau, UZ Tau E, HK Tau, HQ Tau \\
            &       &           &     & J0510+1800,J0423-0120,J0440+2728 & DH Tau \\
            &       &           &     & J0510+1800,J0423-0120,J0422+3058 & BP Tau \\
            &       &           &     & J0510+1800,J0423-0120,J0435+2532 & IP Tau \\
2017/09/02  &   45  & 21-3697   & 1.3 & J0510+1800,J0510+1800,J0426+2327 & DK Tau, GK Tau, V409 Tau, GI Tau,  FT Tau\\
			&       &           &     &  & HO Tau, UZ Tau E, HK Tau, HQ Tau \\
            &       &           &     & J0510+1800,J0510+1800,J0440+2728 & DH Tau \\
            &       &           &     & J0510+1800,J0510+1800,J0422+3058 & BP Tau \\
            &       &           &     & J0510+1800,J0510+1800,J0435+2532 & IP Tau \\
\enddata
\tablecomments{The sample of 32 disks was split into four observing groups. From left to right, Col.~(1) Observing UTC data, Col.~(2) Number of antennas,  Col.~(3) Baseline range,  Col.~(4) Level of precipitable water vapor, Col.~(5) Bandpass, Flux, and Phase calibrator,  Col.~(6) Science targets. }
\tablenotetext{*}{The scheduled phase calibrator (J0426+2327) for these disks was observed at different spectral windows from the science targets, thus phase calibration cannot be applied from the phase calibrator to our targets. We used the weaker check source (J0435+2532) instead to transfer phase solutions.}
\end{deluxetable*}

The mass and age of each source in our sample and in the Taurus disk sample of \citet{andrews2013} are then calculated by comparing the temperature and updated luminosity to \citet{baraffe2015} and non-magnetic \citet{feiden2016} models of pre-main sequence stellar evolution, as in \citet{pascucci2016}.  The combination of both sets of evolutionary tracks cover the full range of spectral types in Taurus disks.  For sources that are more luminous than the youngest isochrone, we choose the youngest 0.5 Myr isochrone and then calculate the stellar mass based on stellar effective temperature. For sources that appear fainter than the main population, we calculate a stellar mass from the isochrone of the average age in the full Taurus sample ($\sim$ 2 Myr). We adopt the stellar dynamical mass measurements from the CO gas rotation for the two spectroscopic binaries (UZ Tau E, \citealt{simon00} and DQ Tau, \citealt{czekala2016}) and two relatively edge-on disks (HN Tau A and HK Tau B, \citealt{simon2017}), all corrected for the {\it Gaia} DR2 distance.

In \citet{lodato2019}, we analyzed the putative population of hidden planets in the subset of sources with substructures.  Most of those host stars have masses measured from gas rotation in the disk \citep{simon00,pietu07,guilloteau14,simon2017}, which should be more accurate than masses estimated from HR diagrams.  The accuracy of host mass was also important to that paper, so that we could compare disk properties to the exoplanet systems around stars of the same mass.  For this paper, the masses are most important as a tool for comparison to the parent sample of Taurus disks, including those disks that were excluded from our sample.  These different goals led to different choices in the method to measure stellar mass.

In Appendix~\ref{sec:source-detail}, we discuss some of the uncertainties in assigning stellar masses and ages to each target.  Although individual stellar masses estimated from evolutionary tracks are marginally consistent with most dynamical measurements, a global comparison indicates that the masses used here are likely underestimated.  The average age of the sample is $\sim 2.3$ Myr, consistent with the approximate age of Taurus, but the age of any individual star is unreliable.



\subsection{Observations}

Our ALMA observations were conducted as program 2016.1.01164.S (PI: Herczeg) in 2017 August--September. The Band 6 receivers were used for all measurements with identical spectral window (SPW) setup. The continuum emission was recorded in two SPWs, which centered at 218 and 233 GHz, each with a bandwidth of 1.875 GHz. The resulting average observing frequency is 225.5 GHz (wavelength of 1.3 mm). 
 Another SPW covered $^{13}$CO and C$^{18}$O $J$=2-1 with a velocity resolution of 0.16 km s$^{-1}$. The remaining SPW was designed to target $^{12}$CO $J$=2-1 line, but was unfortunately incorrectly tuned during the observation. The $^{13}$CO emission were detected in about 1/3 of our sample, which will be presented in a forthcoming paper.  
We adopted the C40-7 antenna configuration to achieve the desired spatial resolution of $\sim 0.\farcs1$.

The selected sample of 32 Taurus disks were split into four different observing groups mainly based on their locations in the sky. One observing group (2017/08/27, see Table \ref{tab:ALMAObservations}) consists of bright disks (mm flux $>50$ mJy obtained from \citealt{andrews2013} and \citealt{akeson2014}) with $\sim 4$ min integration time per source. The other three groups, with mostly faint disks ($<50$ mJy, with exceptions for a few bright disks for observing efficiency), were observed for $\sim 8-9$ min per source. Bandpass and flux calibrators were observed at the beginning of each observing group/block, and a phase calibrator near the science targets was repeatedly recorded every 30--60 s. The observing conditions and calibrators for each observing group are summarized in Table \ref{tab:ALMAObservations}.

Data reduction started with the standard ALMA pipeline calibration, with scripts provided by ALMA staff. This calibration procedure was performed with \textsc{CASA} v4.7.2 for the first observing group (2017/08/18) and v5.1.1 for the later three groups. Following the pipeline, initial phase adjustments were made based on the water vapor radiometer measurements. The standard bandpass, flux, and gain calibrations were then applied accordingly for each measurement set (see Table~\ref{tab:ALMAObservations}). 
In some observations in the second observing group (2017/08/27), the phase calibrator was recorded at different spectral setup from the science targets. We therefore used the weaker check source for phase corrections (see note in Table~\ref{tab:ALMAObservations}).
Self-calibration were performed for our targets, except for the faint GK Tau and HQ Tau, with procedures elaborated in \citet{long2018}. As a result, self-calibration provided visible improvement in image quality that image peak signal-to-noise ratio (SNR) for most disks increased by $\sim 30\%$ and a factor of 2--3 improvement in image SNR was seen for the brightest disks.  After continuum self-calibration, the data visibilities were extracted for further modeling. We then created continuum image for each target using \textit{tclean} with Briggs weighting and a robust parameter of +0.5. Our final continuum images have a typical beam size of $0.\farcs14\times0.\farcs11$ and a median continuum rms of 50 $\mu$Jy beam$^{-1}$ (see peak intensity and noise level for individual disks in Table~\ref{tab:source_prop}).

\begin{figure*}[!t]
\centering
    \includegraphics[width=0.98\textwidth]{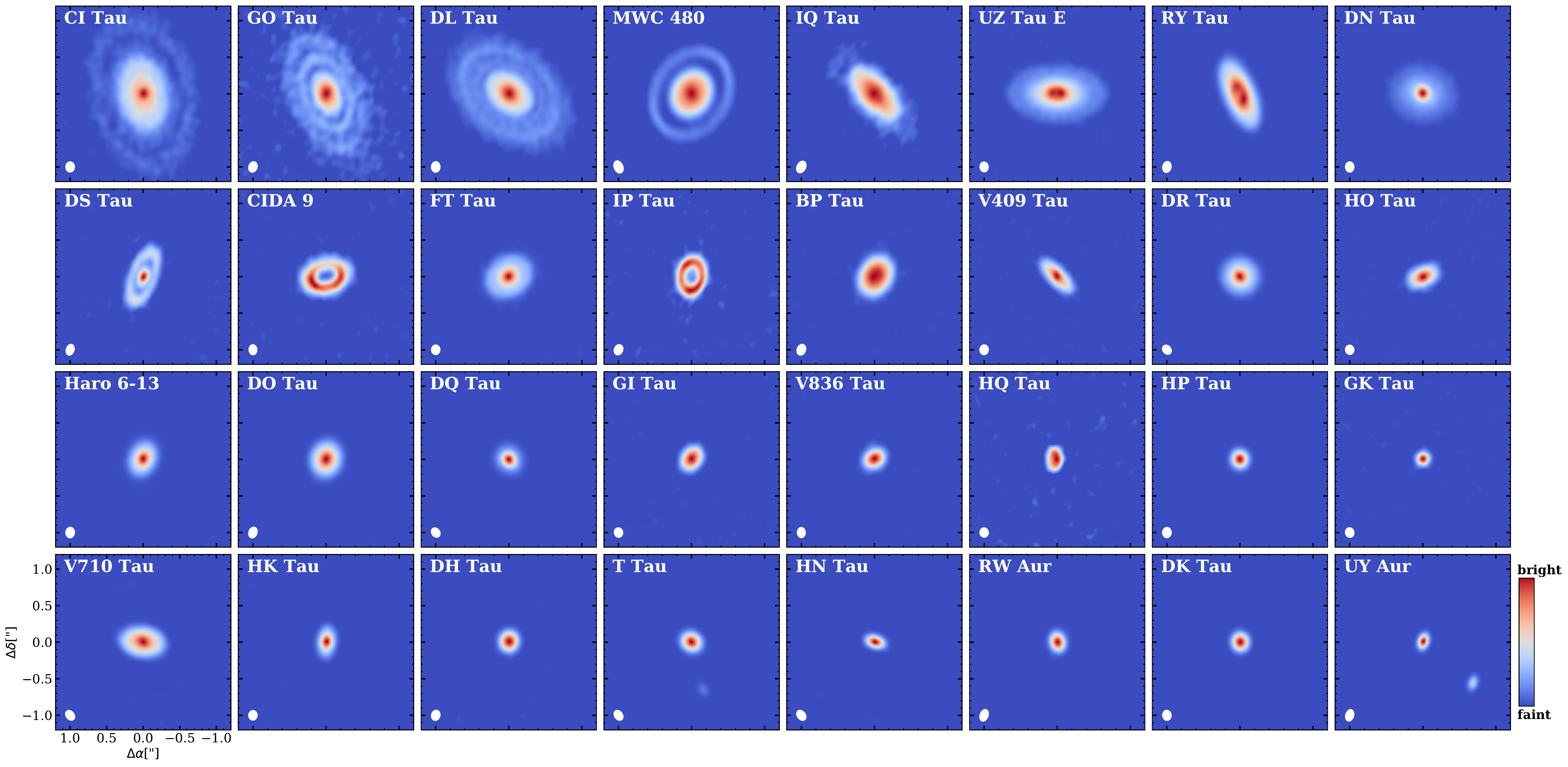} \\ 
    \caption{The 1.3 mm images for our full sample, made with a Briggs weighting of robustness parameter of 0.5. The first 12 panels show images for disks with substructures, followed by the 12 smooth disks around single stars. The last row shows images for the 8 smooth disks in binaries. The images are displayed in order of decreasing disk radii in each sub-sample. To highlight the weak outer emission of a few disks, an asinh scaling function has been applied. Each panel is $2.\farcs4\times2.\farcs4$, with the synthesised beam shown in the left corner. The relative color scale is shown in the right corner. \label{fig:cont_images} }
\end{figure*}

\section{Disk Modeling in the visibility plane} \label{sec:model}

The 32 images of 1.3 mm continuum emission  (Figure~\ref{fig:cont_images}) reveal two types of disks: disks with dust substructures in various numbers, locations, and contrasts, and disks with dust emission peaking in the center and monotonically decreasing outward. The 12 disks with prominent dust gaps and rings have been modeled and discussed in detail in \citet{long2018}. For the other 20 disks, we follow the similar disk modeling procedure in the visibility plane as presented in \citet{long2018} to describe the disk dust distribution. The best-fit models are then used to derive the general disk properties (disk position angle, inclination, mm fluxes and disk radius) for further analysis.

\subsection{Modeling Procedure}
Our model fitting is performed in the visibility plane.  The main procedure is summarized as follows: we first take a model intensity profile and Fourier transform it to create the model visibilities; the fitting is then executed by comparing the model visibilities to data visibilities with the Markov chain Monte Carlo (MCMC) method to derive the best-fit model.

\begin{figure}[!t]
\centering
    \includegraphics[width=0.46\textwidth]{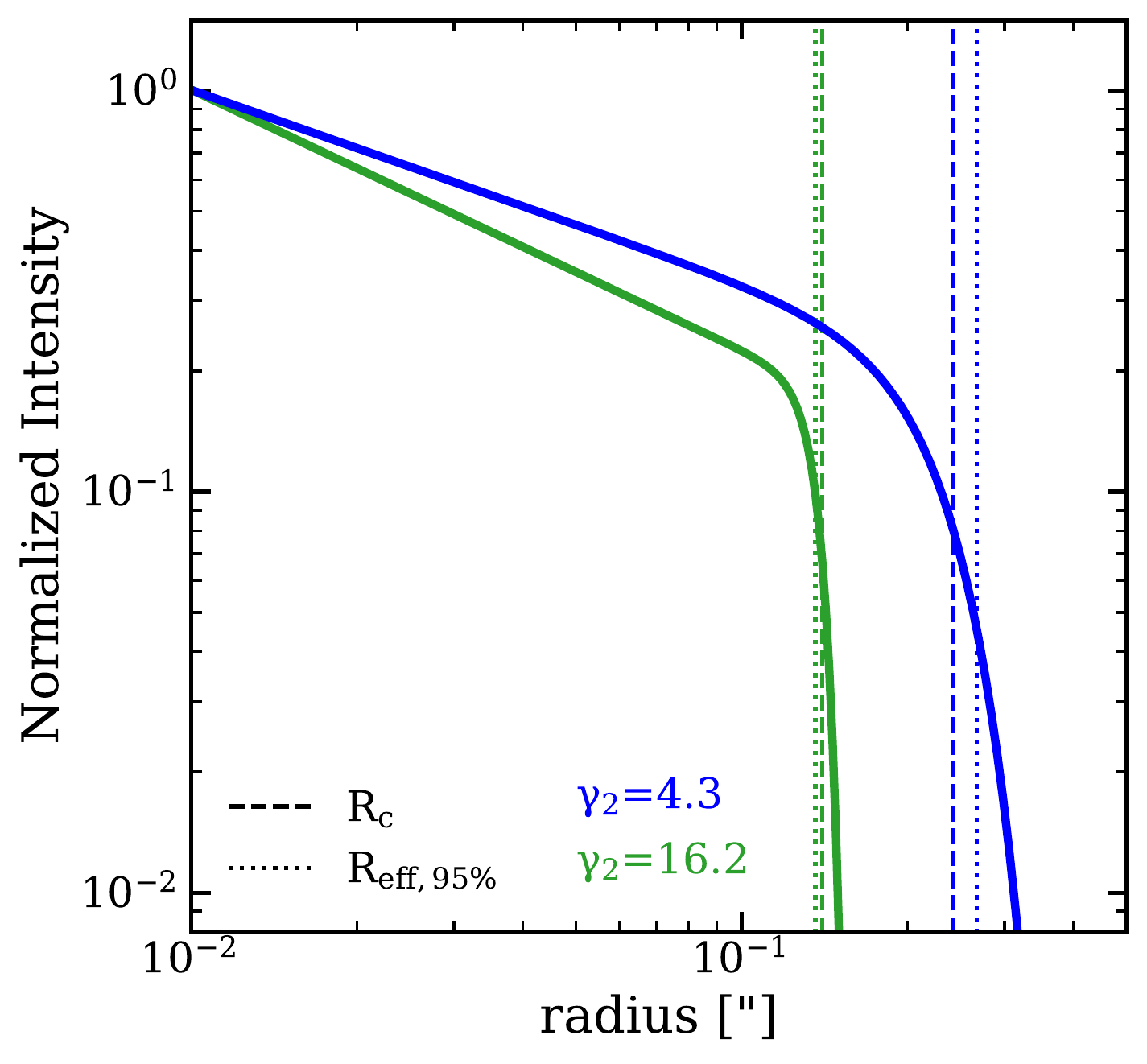} \\
    \caption{Representative profiles of the exponentially tapered power law. The best-fit model profiles for HO Tau (in blue, \{$R_c$, $\gamma_{1}$, $\gamma_{2}$\} = \{0.$\farcs$24, 0.48, 4.3\}) and HN Tau (in green, \{$R_c$, $\gamma_{1}$, $\gamma_{2}$\} = \{0.$\farcs$14, 0.65, 16.2\}) are selected as examples from single stars and binary systems respectively, showing different degree of sharpness of the outer disk. Transition radius ($R_c$) in the model and disk effective radius at 95\% flux encircled for both disks are marked as dashed and dotted lines. 
    \label{fig:model} }
\end{figure}

The choice of model profile is guided by the appearance of the visibility profile. The oscillation pattern in the real part of the visibility profile is seen for a fraction of disks (see also Figure~\ref{fig:model_result_single} in the Appendix), which likely indicates a disk with a sharp outer edge in millimeter dust grains (e.g., \citealt{hogerheijde2016,zhang2016}). We therefore adopt an exponentially tapered power law ($I(R)=A(R/R_{c})^{-\gamma_{1}}\exp[-(R/R_{c})^{\gamma_{2}}]$) as the model intensity profile, in which power law index $\gamma_{1}$ and taper index $\gamma_{2}$ describe the slope of the emission gradient in the inner disk and the sharpness of the falloff beyond the transition radius ($R_c$), respectively (see Figure~\ref{fig:model}). The model is also described by a disk inclination and position angle and phase center offsets. We then apply the \textit{Galario} code \citep{tazzari2018} to Fourier transform the model intensity profile into visibilities sampled with the same uv-coverage. The model visibilities are later compared with data visibilities using \textit{emcee} package \citep{Foreman-Mackey2013}. 
The parameters are explored with 100 walkers and 5000 steps 
for each walker. The burn-in phase for convergency is typically less than 1000 steps. The posterior medians are obtained using the MCMC chains of the last 1000 steps, with the 1$\sigma$ uncertainty for each parameter calculated from 16th and 84th percentiles.

\subsection{Modeling Results}
For single stars in our sample with no detectable substructures, we apply the modeling approach described above to fit the disk dust distribution. For multiple stellar systems (see Table~\ref{tab:source_prop}),  the fitting results are adopted from the companion paper of Manara et al.~(submitted), which fits multiple disk components simultaneously.  Our analysis below only includes the circumprimary disks, which are modeled with the same morphologic function as disks in single stellar systems.

The quality of the best-fit model is checked by inspecting the comparisons of data and model in images, visibility profiles, and radial intensity cuts (see Figure~\ref{fig:model_result_single} in the Appendix). In most cases, the exponentially-tapered power law can well describe the dust emission, with residuals less than 3$\sigma$. For DR Tau and DQ Tau, however, the comparisons of best-fit model to data yield asymmetric residuals of 5-10$\sigma$. The residual in the innermost disk of the spectroscopic binary DQ Tau may be associated with the high orbital eccentricity \citep{mathieu1997}.   
A check for Haro 6-13 also shows 5-10$\sigma$ asymmetric residuals in the inner disk, as well as a possible faint (3$\sigma$) outer disk. 
Since large residuals are seen in all bright disks (high peak SNR), we may miss some fine details and faint substructures that would have been detected with greater sensitivity and spatial resolution \citep[see, e.g.][]{huang2018_sprial}. This is also indicated by the data and model comparison at longer baselines (Figure~\ref{fig:model_result_single}), where our simple model might miss some small-scale structures.
For all disks, the exponentially tapered power law fits better than the Gaussian profile, except for the faint and compact GK Tau where both models work similarly well.

\begin{deluxetable*}{lrcccccrrc}
\tabletypesize{\scriptsize}
\tablecaption{Disk Model Parameters\label{tab:fitting_model}}
\tablewidth{0pt}
\tablehead{
\colhead{Name} & \colhead{$F_\nu$} & \colhead{$R_{\rm eff,68\%}$} & \colhead{$R_{\rm eff,95\%}$} & \colhead{$R_c$} & \colhead{$\gamma_{1}$} & \colhead{$\gamma_{2}$} & \colhead{incl} & \colhead{PA} & \colhead{Source Center} \\
\colhead{} & \colhead{(mJy)} & \colhead{(arcsec)} & \colhead{(arcsec)} & \colhead{(arcsec)} & \colhead{} & \colhead{} &  \colhead{(deg)} & \colhead{(deg)} & \colhead{} 
} 
\colnumbers
\startdata
   CI Tau &  142.40$^{+0.47}_{-0.81}$ &  0.706 &  1.195 & -- &   -- &  -- & 50.0$^{+0.3}_{-0.3}$ &   11.2$^{+0.4}_{-0.4}$ &  04h33m52.03s  +22d50m29.81s \\
 CIDA 9 A &   37.10$^{+0.26}_{-0.20}$ &  0.287 &  0.371 & -- &   -- &  -- & 45.6$^{+0.5}_{-0.5}$ &  102.7$^{+0.7}_{-0.7}$ &  05h05m22.82s  +25d31m30.50s \\
   DL Tau &  170.72$^{+0.93}_{-0.43}$ &  0.702 &  1.033 & -- &   -- &  -- & 45.0$^{+0.2}_{-0.2}$ &   52.1$^{+0.4}_{-0.4}$ &  04h33m39.09s  +25d20m37.79s \\
   DN Tau &   88.61$^{+0.25}_{-0.62}$ &  0.313 &  0.475 & -- &   -- &  -- & 35.2$^{+0.5}_{-0.6}$ &   79.2$^{+1.0}_{-1.0}$ &  04h35m27.39s  +24d14m58.55s \\
   DS Tau &   22.24$^{+0.23}_{-0.23}$ &  0.376 &  0.446 & -- &   -- &  -- & 65.2$^{+0.3}_{-0.3}$ &  159.6$^{+0.4}_{-0.4}$ &  04h47m48.60s  +29d25m10.76s \\
   FT Tau &   89.77$^{+0.27}_{-0.25}$ &  0.264 &  0.357 & -- &   -- &  -- & 35.5$^{+0.4}_{-0.4}$ &  121.8$^{+0.7}_{-0.7}$ &  04h23m39.20s  +24d56m13.86s \\
   GO Tau &   54.76$^{+0.85}_{-0.41}$ &  0.698 &  1.187 & -- &   -- &  -- & 53.9$^{+0.5}_{-0.5}$ &   20.9$^{+0.6}_{-0.6}$ &  04h43m03.08s  +25d20m18.35s \\
   IP Tau &   14.53$^{+0.17}_{-0.18}$ &  0.234 &  0.280 & -- &   -- &  -- & 45.2$^{+0.8}_{-0.9}$ &  173.0$^{+1.1}_{-1.1}$ &  04h24m57.09s  +27d11m56.07s \\
   IQ Tau &   64.11$^{+0.49}_{-0.72}$ &  0.423 &  0.838 & -- &   -- &  -- & 62.1$^{+0.5}_{-0.5}$ &   42.4$^{+0.6}_{-0.6}$ &  04h29m51.57s  +26d06m44.45s \\
  MWC 480 &  267.76$^{+0.51}_{-1.07}$ &  0.345 &  0.878 & -- &   -- &  -- & 36.5$^{+0.2}_{-0.2}$ &  147.5$^{+0.3}_{-0.3}$ &  04h58m46.27s  +29d50m36.51s \\
   RY Tau &  210.40$^{+0.21}_{-0.21}$ &  0.378 &  0.509 & -- &   -- &  -- & 65.0$^{+0.1}_{-0.1}$ &   23.1$^{+0.1}_{-0.1}$ &  04h21m57.42s  +28d26m35.09s \\
 UZ Tau E &  129.52$^{+0.68}_{-0.79}$ &  0.445 &  0.667 & -- &   -- &  -- & 56.1$^{+0.4}_{-0.4}$ &   90.4$^{+0.4}_{-0.4}$ &  04h32m43.08s  +25d52m30.63s \\
 \hline
    BP Tau &  45.15$^{+0.19}_{-0.14}$ & 0.226 &  0.321 &  0.273 &  0.10$^{+0.03}_{-0.03}$  &  3.93$^{+0.24}_{- 0.24}$ & 38.2$^{+0.5}_{-0.5}$  & 151.1$^{+1.0}_{-1.0}$  & 04h19m15.85s  +29d06m26.48s \\
    DO Tau & 123.76$^{+0.17}_{-0.27}$ & 0.183 &  0.263 &  0.247 &  0.53$^{+0.00}_{-0.00}$  &  4.97$^{+0.14}_{- 0.14}$ & 27.6$^{+0.3}_{-0.3}$  & 170.0$^{+0.9}_{-0.9}$  & 04h38m28.60s  +26d10m49.08s \\
    DQ Tau &  69.27$^{+0.15}_{-0.19}$ & 0.124 &  0.219 &  0.166 &  0.80$^{+0.03}_{-0.03}$  &  2.37$^{+0.12}_{- 0.12}$ & 16.1$^{+1.2}_{-1.2}$ &  20.3$^{+4.3}_{-4.3}$ & 04h46m53.06s  +16d59m59.89s \\
    DR Tau & 127.18$^{+0.20}_{-0.22}$ & 0.188 &  0.276 &  0.267 &  0.70$^{+0.00}_{-0.00}$  &  5.37$^{+0.16}_{- 0.16}$ &  5.4$^{+2.1}_{-2.6}$  &   3.4$^{+8.2}_{-8.0}$  &  04h47m06.22s  +16d58m42.55s \\
    GI Tau &  17.69$^{+0.25}_{-0.07}$ & 0.145 &  0.190 &  0.193 &  0.39$^{+0.05}_{-0.05}$  &  9.69$^{+5.56}_{- 3.66}$ & 43.8$^{+1.1}_{-1.1}$ & 143.7$^{+1.9}_{-1.6}$  & 04h33m34.07s  +24d21m16.70s \\
    GK Tau &   5.15$^{+0.19}_{-0.11}$ & 0.065 &  0.099 &  0.085 &  0.53$^{+0.59}_{-0.91}$  &  3.47$^{+8.64}_{- 3.25}$ & 40.2$^{+5.9}_{-6.2}$  & 119.9$^{+8.9}_{-9.1}$ & 04h33m34.57s  +24d21m05.49s \\
 Haro 6-13 & 137.10$^{+0.24}_{-0.21}$ & 0.185 &  0.264 &  0.268 &  0.78$^{+0.00}_{-0.00}$  &  7.25$^{+0.32}_{- 0.32}$ & 41.1$^{+0.3}_{-0.3}$  & 154.2$^{+0.3}_{-0.3}$  & 04h32m15.42s  +24d28m59.21s \\
    HO Tau &  17.72$^{+0.20}_{-0.17}$ & 0.183 &  0.267 &  0.242 &  0.48$^{+0.05}_{-0.05}$  &  4.30$^{+0.76}_{- 0.65}$ & 55.0$^{+0.8}_{-0.8}$  & 116.3$^{+1.0}_{-1.0}$ & 04h35m20.22s  +22d32m14.27s \\
    HP Tau &  49.33$^{+0.16}_{-0.15}$ & 0.090 &  0.125 &  0.127 &  0.68$^{+0.06}_{-0.06}$  &  8.31$^{+3.12}_{- 2.45}$ & 18.3$^{+1.2}_{-1.4}$  &  56.5$^{+4.6}_{-4.3}$  & 04h35m52.79s  +22d54m22.93s \\
    HQ Tau &   3.98$^{+0.08}_{-0.17}$ & 0.129 &  0.155 &  0.158 & -0.21$^{+0.29}_{-0.34}$  & 16.40$^{+6.89}_{-11.51}$ & 53.8$^{+3.2}_{-3.2}$  & 179.1$^{+3.2}_{-3.4}$  & 04h35m47.35s  +22d50m21.36s \\
  V409 Tau &  20.22$^{+0.12}_{-0.18}$ & 0.239 &  0.311 &  0.324 &  0.59$^{+0.03}_{-0.03}$  & 16.11$^{+6.25}_{- 5.98}$ & 69.3$^{+0.3}_{-0.3}$  &  44.8$^{+0.5}_{-0.5}$  & 04h18m10.79s  +25d19m56.97s \\
  V836 Tau &  26.24$^{+0.16}_{-0.12}$ & 0.128 &  0.188 &  0.156 &  0.22$^{+0.08}_{-0.10}$  &  3.52$^{+0.55}_{- 0.52}$ & 43.1$^{+0.8}_{-0.8}$  & 117.6$^{+1.3}_{-1.3}$  & 05h03m06.60s  +25d23m19.29s \\
  \hline
    DH Tau A&  26.68$^{+0.13}_{-0.12}$ & 0.105 &  0.146 &  0.140 &  0.38$^{+0.07}_{-0.07}$  &  5.73$^{+ 1.35}_{- 1.08}$ & 16.9$^{+2.0}_{-2.2}$  &  18.8$^{+ 7.1}_{- 7.2} $ & 04h29m41.56s  +26d32m57.76s \\
    DK Tau A&  30.08$^{+0.14}_{-0.09}$ & 0.092 &  0.117 &  0.120 &  0.60$^{+0.03}_{-0.03}$  & 38.93$^{+14.57}_{-20.79}$ & 12.8$^{+2.5}_{-2.8}$  &   4.4$^{+10.1}_{- 9.4}$  & 04h30m44.25s  +26d01m24.35s \\
    HK Tau A&  33.15$^{+0.15}_{-0.13}$ & 0.156 &  0.216 &  0.230 &  0.92$^{+0.01}_{-0.01}$  & 21.36$^{+17.75}_{-10.06}$ & 56.9$^{+0.5}_{-0.5}$  & 174.9$^{+ 0.5}_{- 0.5}$  & 04h31m50.58s  +24d24m17.37s \\
    HN Tau A&  12.30$^{+0.12}_{-0.18}$ & 0.104 &  0.136 &  0.140 &  0.65$^{+0.05}_{-0.05}$  & 16.19$^{+ 4.74}_{- 7.31}$ & 69.8$^{+1.4}_{-1.3}$  &  85.3$^{+ 0.7}_{- 0.6}$  & 04h33m39.38s  +17d51m51.98s \\
    RW Aur A&  35.60$^{+0.28}_{-0.27}$ & 0.101 &  0.132 &  0.140 &  0.70$^{+0.02}_{-0.02}$  & 26.24$^{+14.96}_{-12.61}$ & 55.1$^{+0.5}_{-0.4}$  &  41.1$^{+ 0.6}_{- 0.6}$  & 05h07m49.57s  +30d24m04.70s \\
     T Tau N& 179.72$^{+0.22}_{-0.22}$ & 0.111 &  0.143 &  0.150 &  0.68$^{+0.00}_{-0.00}$  & 49.58$^{+ 0.78}_{- 1.75}$ & 28.2$^{+0.2}_{-0.2}$  &  87.5$^{+ 0.5}_{- 0.5}$  & 04h21m59.45s  +19d32m06.18s \\
    UY Aur A&  19.96$^{+1.07}_{-1.06}$ & 0.033 &  0.044 &  0.040 &  0.24$^{+0.97}_{-2.05}$  &  7.10$^{+12.59}_{- 5.55}$ & 23.5$^{+7.8}_{-6.6}$  & 125.7$^{+10.3}_{-10.9}$  & 04h51m47.40s  +30d47m13.10s \\
  V710 Tau A&  55.20$^{+0.19}_{-0.14}$ & 0.238 &  0.317 &  0.320 &  0.48$^{+0.01}_{-0.01}$  &  8.82$^{+ 0.62}_{- 0.59}$ & 48.9$^{+0.3}_{-0.3}$  &  84.3$^{+ 0.4}_{- 0.4}$  & 04h31m57.81s  +18d21m37.64s \\
\enddata
\tablecomments{The power law index $\gamma_{1}$ and taper index $\gamma_{2}$, as well as the disk inclination and PA are parameters fitted with MCMC. Total flux ($F_\nu$) and effective radius ($R_{\rm eff}$, with both 68\% and 95\% flux encircled) are derived from the best-fit intensity profile for each disk. The quoted uncertainties are the interval from the 16th to the 84th percentile of the model chains and scaled by the square root of the reduced $\chi^2$ of the fit. Uncertainties for all radii are extremely small (at a level of 0.$\farcs$002) and thus not showing. The source center is derived by applying the fitted phase center offsets to the image center.}
\end{deluxetable*}

\subsubsection{Best-fit profile parameters}
The best-fit model parameters, including power-law and taper indices, inclination, and position angle, are summarized in Table~\ref{tab:fitting_model}.  The taper index $\gamma_{2}$ describes the profile of the outer disk (Figure~\ref{fig:model}).   The taper index is generally higher than those of the widely-used similarity solution, implying sharp outer edges of dust disks. Most of the disks in binary systems have the sharpest outer disk edges (larger $\gamma_{2}$ index) in our sample, hinting for higher level of outer disk truncation by close companions (see the detailed discussion in Manara et al.~submitted).

The distribution of materials in the inner disk is characterized by the power law index $\gamma_{1}$. The negative $\gamma_{1}$ index of HQ Tau indicates depletion towards the inner region,  perhaps indicating the existence of a dust cavity that is not well resolved in our current data. Except for HQ Tau, most smooth disks in our sample have similar inner disk profiles, with the median $\gamma_{1}$ value of 0.56 and a standard deviation of 0.26. BP Tau has a peculiar flat inner disk, with $\gamma_{1}$ of only 0.1.

The listed uncertainties for the fitted parameters are adopted as the 16th to 84th percentile range of the posterior distribution for each parameter, and are then scaled by the square root of the reduced $\chi^2$ of the fit. These uncertainties correspond to statistical uncertainties and are likely underestimated. 

Since some targets were observed in multiple nights and with different beam shapes, differences between separate fits to the sets of observations provide us with an independent estimate for the observational errors. We include the fitting results for a few disks in Table~\ref{tab:fitting_model_multiple} in the Appendix. These fits demonstrate that the inclinations and position angles have a precision of $\sim 1-2$ deg, and the effective radii are precise to $\sim 3$\%, fluxes to 5\%.  The power-law and taper indices have larger uncertainties, although the values are generally similar.  The scale of the uncertainty depends on disk brightness and disk size.  The two disks without self-calibration, GK Tau and HQ Tau, have larger uncertainties derived from the fitting than the average, likely due both to their faintness and their compactness.


\subsubsection{Fluxes and Sizes of Dust Disks}

We summarize the disk mm fluxes and disk sizes in Table~\ref{tab:fitting_model}. Based on the best-fit model profiles, the disk mm flux densities and dust disk sizes are derived as in \citet{long2018}. The mm continuum fluxes for each disk, measured by integrating over the intensity profile, are broadly consistent with pre-ALMA flux measurements \citep{andrews2013}, if taking into account a 10--15\% systematic uncertainty. 

The dust disk size is defined here as the radius that encircles some fraction of the total flux, calculated for 68\% and 95\% for direct comparisons with previous studies.  
For disks with a sharp outer edge (large $\gamma_2$), the disk $R_{\rm eff,95\%}$ almost overlaps with $R_c$, the transition radius of the power law model profile, while for disks with shallower variations, $R_{\rm eff,95\%}$ is typically 10-20\% further out than $R_c$. For the 11 disks in our sample with $R_{\rm eff,68\%}$ measured by \citet{tripathi2017} with SMA observations at 0.88 mm ($\sim$340 GHz), the disk radii at 0.88 mm are systematically larger than our measurements at 1.3 mm by an average factor of 1.6. The largest differences are seen for DK Tau, Haro 6-13, and HP Tau, which are all more than two times larger at 0.88 mm than measured here.  These three disks are smooth and lack substructures in our observations, and are compact enough that the 0.88 mm measurements may be affected by the lower angular resolution of SMA (typical resolution of $0.\farcs5$).
Though the continuum emission at longer wavelength is expected to be more compact as a consequence of dust grain growth and radial drift (e.g., \citealt{perez2012,perez2015}, \citealt{menu2014}, \citealt{tazzari2016}), when the gas pressure profile is smooth in the outer disk, a factor of 2 difference at such close wavelengths (grain sizes) is hard to be produced in dust evolution models.

\begin{figure*}[!t]
\centering
    \includegraphics[width=0.48\textwidth]{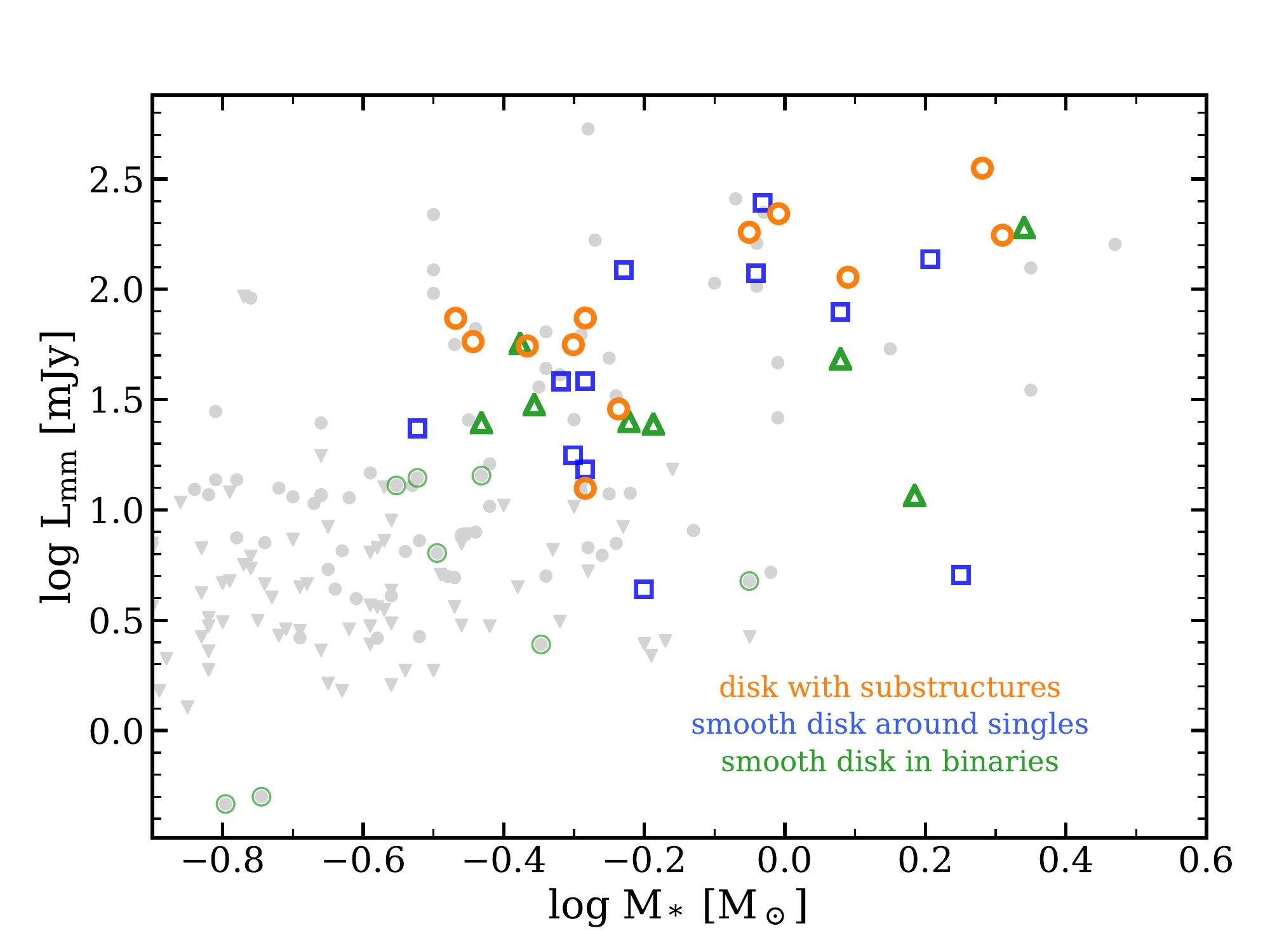}
    \includegraphics[width=0.48\textwidth]{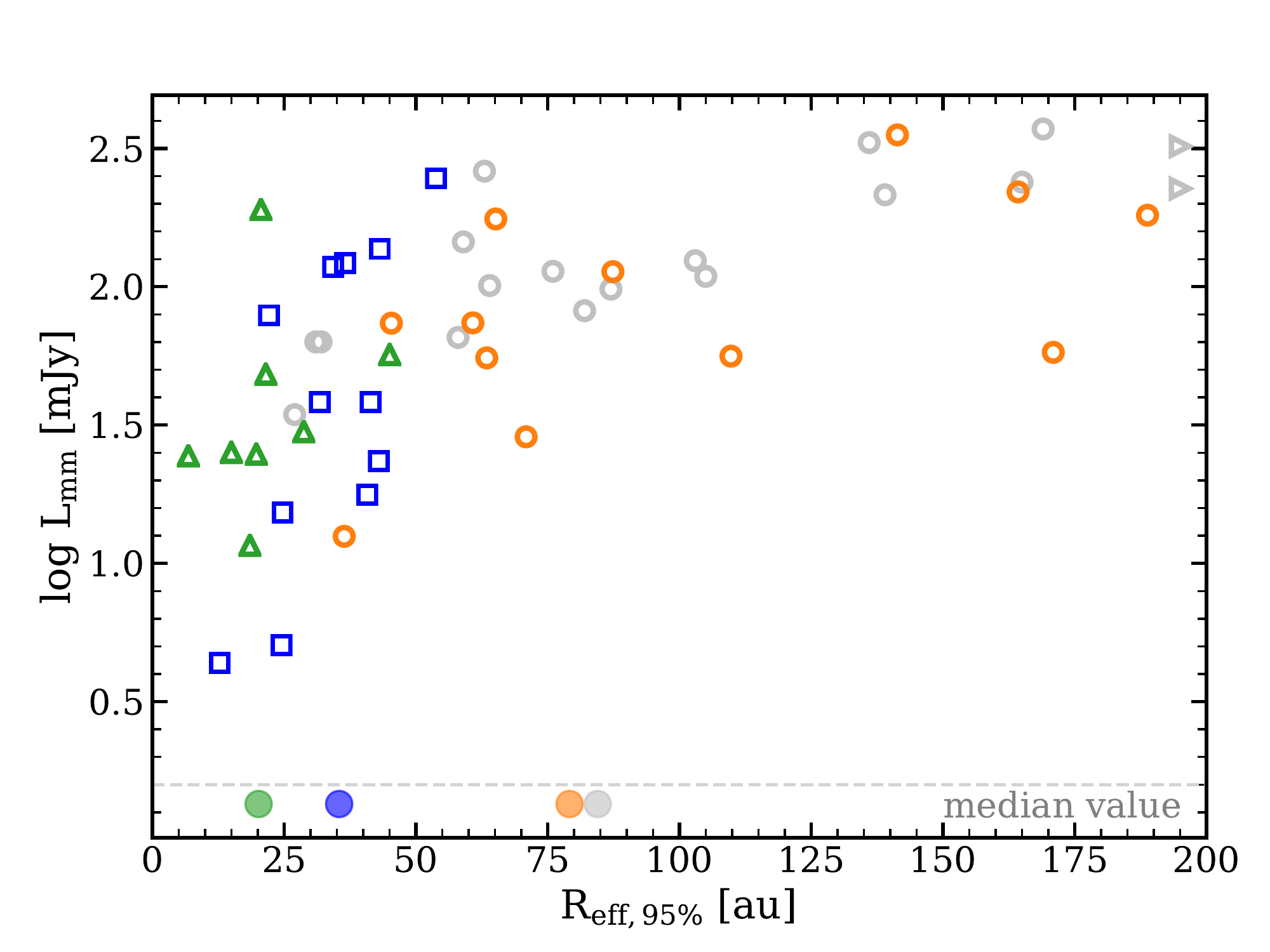} \\
    \caption{Left: stellar mass vs. disk continuum luminosity (scaled to 140 pc) for the 12 disks with substructures (in orange), the 12 smooth disks in singles (in blue), and the 8 smooth disks in binaries/multiples (in green). The other Taurus members (in grey, upper limits in triangles) of \citet{andrews2013} are shown as background comparison, updated for measurements of the secondary disks in our binary sample (in light green circles, plus two disks of UZ Tau Wab). Right: disk effective radius ($R_{\rm eff,95\%}$) vs. disk continuum luminosity for the same color notation. The dots represent the median disk radii in the three sub-samples. The DSHARP sample is included (open grey circles) for a direct comparison, whose disk outer radii are also adopted as 95\% flux encircled. The two largest disks in DSHARP extending to 250 au are marked as right-handed triangles. 
    \label{fig:ms_md_r}}
\end{figure*}

\section{Results} \label{sec:result}

\subsection{Disk sub-sample Category} \label{sec: category}
Our high-resolution ALMA Survey consists of 32 disks in the Taurus Clouds, one of the largest samples studied at $0\farcs1$ resolution. In \citet{long2018}, we described the 12 disks that shows prominent disk substructures mainly based on the inspection of radial intensity profiles, for which dust emission could not be fit with a single central component.  An exponentially tapered power law provides a good fit to most of the other 20 disks (see Section~\ref{sec:model}), which confirms the robustness of our previous selection in \citet{long2018}.  
These 20 disks are therefore referred as smooth disks for their lack of resolved structures, although these disks might host small-scale substructures that are not yet identified in our data.
The 20 smooth disks are further separated into 12 disks around single stars and 8 disks in binary (or multiple) systems that have separations in the range of $0\farcs7-3\farcs5$ and may be affected by tidal interactions
(e.g., \citealt{artymowicz1994, harris2012, long2018-chaI}). Based on the dust morphology (and the effect of stellar multiplicity on dust distribution), this division leads to three catagories of disks (see also Table~\ref{tab:source_prop}): 

\textit{Disks with substructures:} 12 disks show remarkable dust substructures, including four disks with inner dust cavities (plus additional rings in two disks), three disks with inner disk encircled by a single ring, and five disks with inner disk encircled by multiple rings. The inner disk is modeled by either a Gaussian profile or an exponentially tapered power law, and each substructure component is modeled by a Gaussian ring to infer to gap and ring properties. The possible formation mechanisms for disk substructures are discussed in \citet{long2018} and \citet{lodato2019} based on the derived gap and ring properties. Two multiple systems, CIDA 9 (separation of $2\farcs34$) and UZ Tau E (separation of $3\farcs56$ from the close binary UZ Tau Wab) are included in this sub-sample. 

\textit{Smooth disks around singles:} 12 disks around stars in single stellar systems are well described by one model component and do not show apparent substructures at current resolution. Some 5--10$\sigma$ residuals are seen in three bright disks (DR Tau, Haro 6-13, and DQ Tau), which may host unresolved fine substructures in the inner disks. The spectroscopic binary DQ Tau (separation of $<$0.1 au) is included in this sub-sample, since the inner cavity caused by the binary motion remains unresolved in our data. The possibly negative power law index in the very faint HQ Tau may also suggest dust depletion in the inner disk.


\textit{Smooth disks in binaries/multiples:} 8 disks around primary stars in multiple stellar systems that appear smooth in our observations. The disks around the additional stellar components are detected in all but two systems (DH Tau and V710 Tau). A detailed discussion about this sub-sample is presented in Manara et al.~(submitted).

\begin{figure}[!t]
\centering
    \includegraphics[width=0.45\textwidth]{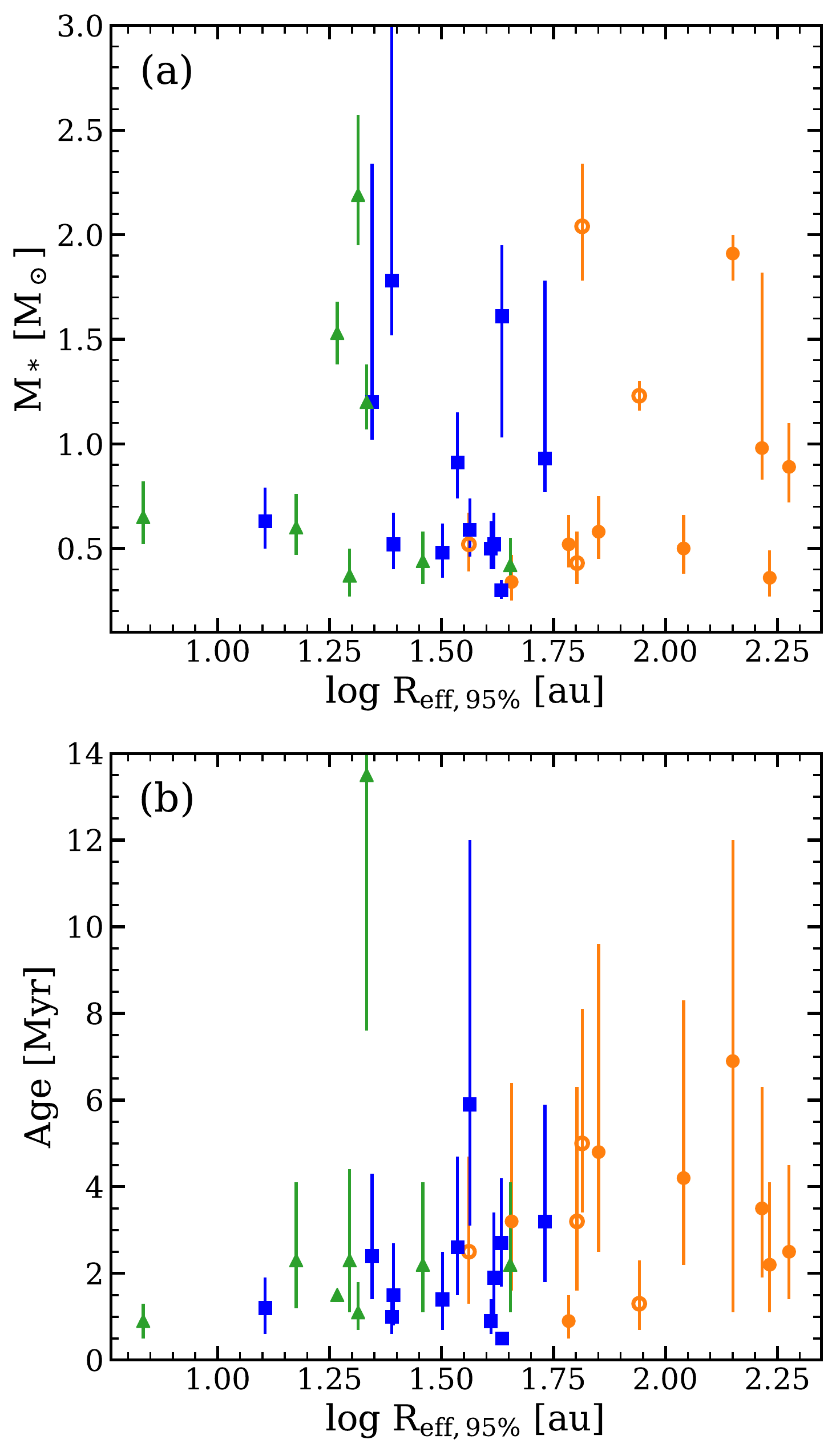} \\
    \caption{stellar mass (a) and stellar age (b) comparison for three sub-samples. Disk dust radii are chosen as x-axis to separate the sub-samples (orange: disks with substructures and open circles for disks with inner cavities, blue: smooth disks around singles, green: smooth disks in binaries). \label{fig:hist} }
\end{figure}

\subsection{Comparisons of stellar and disk properties in the three sub-samples}

In this section, we will assess the similarities and diversities in stellar mass, disk brightness, system age, disk radius and dust profile for our defined sub-samples (Section~\ref{sec: category}), to evaluate the general properties for systems with detectable substructures.

\subsubsection{Comparison of stellar masses}

Our ALMA sample covers a wide range in stellar mass, from $\sim$0.3~M$_\odot$ (set by the prior selection for stars with SpTy earlier than M3) to $\sim2$~M$_\odot$, but populates in the early M and late K type stars. Stellar masses in each of the three sub-samples span the full range of our whole sample, as seen in Figure~\ref{fig:ms_md_r}. By performing the two-sample KS test using ks\_2samp task in \textsc{Python scipy} package, we find that stellar mass distribution in disks with substructures is indistinguishable from that of the smooth disks ($p=94\%$, or from that of the smooth disks in singles with $p=98\%$). The similar stellar mass distribution in the three sub-samples is also evident in Figure~\ref{fig:hist}, with most disks clustered around 0.5 M$_\odot$ and a few disks reaching beyond 1 M$_\odot$ in all three sub-samples.

\subsubsection{Comparison of disk continuum luminosities}

We adopt here the continuum luminosity, $L_{\rm mm} = F_\nu(d/140)^2$, where $d$ is the \textit{Gaia} DR2 distance for individual disks, to present the disk brightness. This quantity is directly proportional to the commonly computed disk dust mass, when assuming uniform dust temperature and dust opacity in all disks.

The disk millimeter luminosity in our full sample spans almost two orders of magnitude (see Figure~\ref{fig:ms_md_r}), from merely 4 mJy to $>300$ mJy, with a median luminosity of $\sim55$ mJy. The set of disks with substructures is slightly brighter than the smooth disk sample, with average disk luminosity a factor of $\sim$2 higher than that of smooth disks in single stars and a factor of $\sim$3 than that of the binary sample. 
Our KS tests suggest that the continuum luminosity distributions for the disks with substructures and the smooth disks in singles are not drawn from different parent samples ($p=18\%$), while clear difference is seen from the comparison with the smooth disks in binaries ($p=4\%$). 
A fraction of smooth disks have comparable brightness as the disks with substructures but distinct smaller disk radii seen at millimeter dust grains (the right panel of Figure~\ref{fig:ms_md_r}).
In the stellar mass range of 0.3--1.0 $M_{\odot}$, our selected sample is still highly underrepresented in the fainter disk population as seen from the full Taurus sample. These faint disks include many close binaries and sources with high extinction, which were left out from our initial selection criterion (see Appendix~\ref{sec:source-selection}).

\subsubsection{Comparison of stellar ages}
Our selected disks have a median age of $\sim 2.3$ Myr, representative of the whole Taurus region. Disks with substructures appear older with a large spread in ages (Figure~\ref{fig:hist}). The median age for disks with substructures is about 3.2 Myr, slightly older than that of the smooth disk sample of 2 Myr (Figure~\ref{fig:HRD}).
However, this age difference is not statistically significant between disks with substructures and smooth disks in singles, in which a two-sample KS test returns a P-value of 15\%. The age distribution indeed looks different when comparing the disks with substructures with smooth disks in binaries ($p=2\%$). As seen in Figure~\ref{fig:spatial}, the full sample is well-mixed in spatial distribution, mostly along the edge of the main filaments. No apparent large age difference emerges from the sample spatial distribution.  These comparisons are also challenging because of the uncertainties in measuring ages \citep[e.g.][]{soderblom14}.

\begin{figure*}[!t]
\centering
    \includegraphics[width=0.98\textwidth]{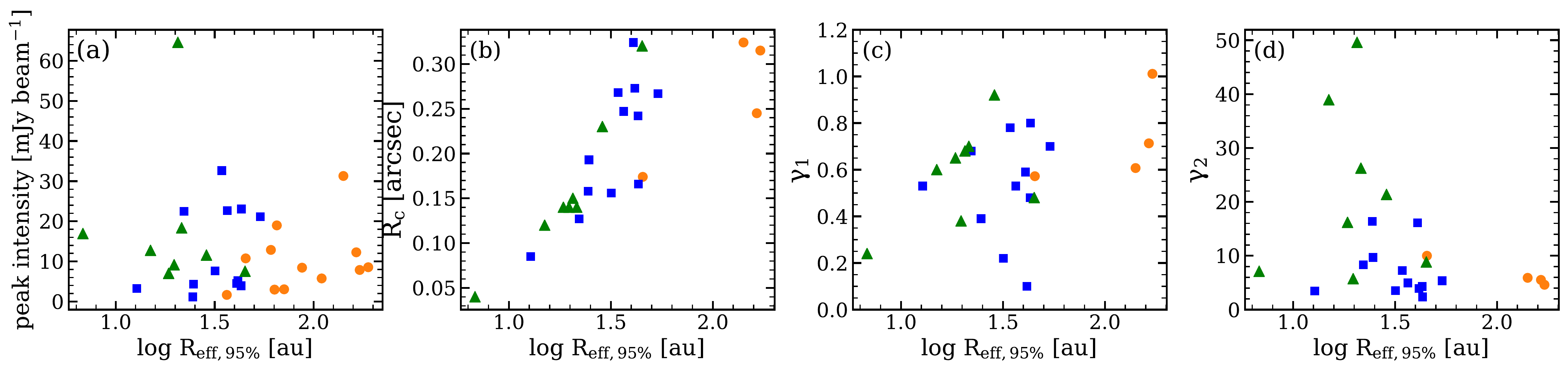} \\ 
    \caption{Inner disk core comparison for three sub-samples (orange for disks with substructures, blue for smooth disks in singles and green for smooth disks in binaries): (a) peak brightness, (b) transition radius $R_{\rm c}$, (c) power law index $\gamma_1$, (d) taper index $\gamma_2$. For the last three panels, only disks modeled with the tapered power law profile are included. 
    \label{fig:inner_core} }
\end{figure*}

\subsubsection{Comparison of dust disk sizes}

The most prominent difference between smooth disks and substructured disks is seen in the size of dust emission, hereafter measured as the effective radius that encircles 95\% of the total flux ($R_{\rm eff,95\%}$). The general results also hold when choosing $R_{\rm eff,68\%}$ as our disk radius definition, since both metrics take into account the outer rings in most cases.

Disks with substructures have continuum emission radii that range from 40 to 200 au, while the smooth disk sample all have radii $\lesssim 55$ au, $\sim 80\%$ of which are between 20--40 au. In other words, disks with effective radius larger than 55 au all show gaps and rings in our sample. The disk size difference is clearly visible in Figure~\ref{fig:ms_md_r} for the three sub-samples, in which disks with substructures have typical dust disk size larger than the smooth disk sample (i.e. a factor of 2--3 larger in median sizes). 
IP Tau, the disk with inner cavity, and FT Tau, the disk with low-contrast emission bump, are the smallest disks in the substructure sample, and with sizes comparable to these of the larger end of the smooth disks.

In addition, the smooth disks in binaries are generally more compact than those in single systems, which likely results from the tidal interaction in binary systems (e.g., \citealt{artymowicz1994,miranda2015}). Most disks in the binary sample have sizes smaller than 30 au. The V710 Tau A disk is the most extended disk ($R_{\rm eff,95\%}\sim 45$ au) in our binary sample; in this system the southern component is not detected in our ALMA observations.  

\subsubsection{Comparison of disk dust profiles}

As established in the previous subsection, disks with substructures are generally more extended in our sample. In this subsection, we demonstrate that these larger radii are obtained because of the presence of outer rings. As seen in Figure~\ref{fig:cont_images}, the inner emission cores for some of the extended ring disks actually have similar extents to the smooth disks.  
Meanwhile, peak brightness distributions are indistinguishable among the three sub-samples, though the T Tau N disk is extremely bright (see Figure~\ref{fig:inner_core}).

We further explore the disk profiles for the inner emission cores in extended ring disks and compact smooth disks. 
In \citet{long2018}, we have employed models with the fewest number of parameters to describe the dust emission morphologies, therefore the inner cores of some disks were modeled with Gaussian profiles. 
In the comparison of disk profile parameters, we thus only include the four disks that were modeled with the tapered power law profile for their inner emission cores when a Gaussian profile could not work equally well. As seen in Figure~\ref{fig:inner_core}, the inner cores of ring disks resemble the smooth disks, with values of disk transition radius ($R_{\rm c}$), power law index ($\gamma_1$), and taper index ($\gamma_2$) well within the parameter ranges of the smooth disk sample. Another four disks with inner cores modeled with Gaussian profiles also have small sizes, with Gaussian radius less than $0\farcs2$. 
The inner cores of ring disks have similar steep outer edge to smooth disks around single stars, while in general shallower than those in binaries. 

\section{Discussion} \label{sec:discussion}

\subsection{The appearance of disk substructures} 
Disk substructures are present in disks across a wide range of parameter space. In our Taurus sample, we detect disk substructures in all spectral types from A to M3 (the hard cut of our sample selection). 
A similar spectral type coverage is found within the DSHARP sample (the 18 disks with annular substructures, \citealt{huang2018_ring}), with disks mainly selected from the Lupus and Ophiuchus star-forming region \citep{andrews2018_dsharp}. Most of the disk substructures are revealed from systems of early-type stars (see also a recent compilation of 16 disks with multi-rings by \citealt{vandermarel2019}), because 1) any serendipitous discovery more likely comes from the preferentially targeted brighter disks, which are linked to earlier spectral types (the stellar mass--disk mass scaling relation, e.g., \citealt{pascucci2016}), 2) specialized substructure surveys (e.g., DSHARP) also selected brighter disks to achieve a sensitivity/contrast criterion \citep{andrews2018_dsharp}, 3) our survey, though covering fainter disks, only probes down to M3 stars.

The existing observations are largely biased towards brighter disks; even our survey, which is designed to include as broad as the range in disk brightness, implies a high occurrence rate of disk substructures among bright disks. Most of the faint disks in our sample have small disk radius (typical $R_{\rm eff,95\%}\sim30$ au, see Figure~\ref{fig:ms_md_r}), in which substructures may not be captured by our $\sim$15 au beam. Given the current observational biases and the observed disk luminosity--size relation \citep{tripathi2017,tazzari2017}, higher resolution ALMA observations for M dwarfs (or even brown dwarfs) and faint disks are needed to build a more complete picture of disk substructures.

Dust rings are detected in both young embedded sources (e.g., HL Tau, \citealt{alma2015}; GY 91, \citealt{shenhan2018}) and more evolved disks (e.g., TW Hydra, \citealt{andrews2016}). In our Taurus sample, substructures are found in systems across an age range of 1--6 Myr (see Figure~\ref{fig:hist}).
Even though the age of individual sources remains poorly determined, the wide age difference is still informative. Disk substructures likely form at a very early stage \citep[e.g.][]{alma2015,shenhan2018} and are sustained in some way for at least a few Myr, although at least one Class I disk, that of TMC1A, is smooth at a resolution of $\sim$8 au \citep{harsono2018}. Current studies have not yet come to firm conclusions about the origin of disk substructures, as a diverse set of mechanisms are capable of reproducing the observed disk patterns. Analyses show no obvious trend between stellar luminosities and the gap/ring locations \citep{long2018,huang2018_ring,vandermarel2019}, thus discarding snow lines as the universal mechanism for disk gap and ring formation. Though no secure evidence has been found to support hidden planets as the cause of gaps in disks \citep{testi2015,guidi2018}, it remains a promising and intriguing explanation, while it opens the question of how relative high mass planets ( $\gtrsim$ Neptune-mass) can form at early disk ages, especially at large separations ($> 50$ au). The assembly of planets may be rapid and happens very early on. The Class I disks might be the key for exploring the onset of disk morphological transition and towards the first steps of planet formation.

Our disks with substructures have similar radial extents as the DSHARP sample (see $R_{\rm eff,95\%}$ comparison in Figure~\ref{fig:ms_md_r}), from $\sim$30 au to $\sim$200 au. The selection criteria of the DSHARP sample inevitably lead to a strong bias towards larger disks \citep{andrews2018_dsharp}. Our blind search of disk substructures in a sample with diverse brightness (also diverse dust disk radius as expected from disk luminosity--size relation), however, results in a preference of finding disk substructures in larger disks (regardless of disk brightness). 
A recent study of 16 multi-ring disks compiled from literature by \citet{vandermarel2019} suggests that the average disk outer radius for the 12 younger disks is a factor of two larger than that of the 4 oldest systems. This trend is not seen in both our Taurus sample and the DSHARP sample, as many young disks have a small radius and the oldest disks (e.g., MWC 480 in our sample) are relatively extended. The small number of older systems observed so far prevents us from drawing any final conclusion. 


\begin{figure*}[!t]
\centering
    \includegraphics[width=0.98\textwidth]{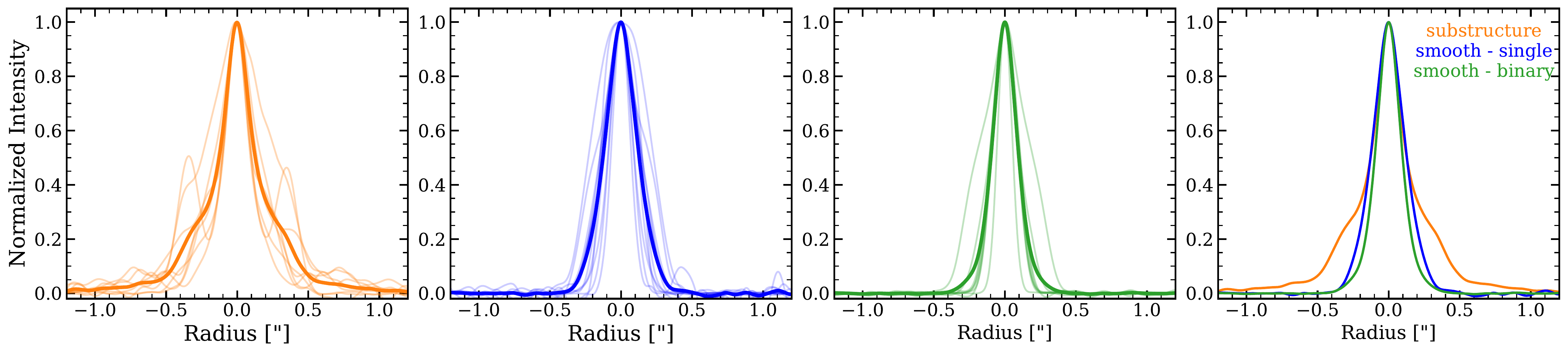} \\ 
    \caption{The comparison of radial profiles, extracted along the disk major axis in the image from left to right for the disks with substructures (excluding the two disks with larger inner cavities), smooth disks around singles, and smooth disks in binaries. The normalized radial profiles for individual disks are shown in light color and the average profiles are shown in thick lines. A straightforward comparison of the three average profiles is drawn in the rightmost panel.
    \label{fig:cont_rp} }
\end{figure*}

\subsection{Disk substructures in compact disks}

Spatially extended disks in our sample show gaps and rings, with diversity in the number and location of the rings and their contrast with gaps.  The smaller disks, 
however, appear smooth in their radial brightness profiles (see Figure~\ref{fig:cont_rp}). This raises the questions of whether our observations are missing some very faint rings at large radii, and whether smaller disks are scaled-down versions of substructures seen in the larger disks.

The comparison of the average disk radial profiles in our defined sub-samples (Figure~\ref{fig:cont_rp}) shows that 1) the inner 0.$\farcs$25 emission core for disks with substructures overlaps with the average profile of the smooth disks in single stars; 2) broader emission appearing as a shoulder spanning from 0.$\farcs$3 to 0.$\farcs$5 followed by a shallow wing extending to 1$\farcs$ is seen in the sample with substructures; 3) disks in binary systems are more compact overall. 
Some rings in the outer disks are very faint (3-10$\sigma$), seen as the wing in the average profile. Given the nearly uniform noise level in the images and similar peak brightness distributions (see Figure~\ref{fig:inner_core}), substructures with similar/stronger significance (i.e. brightness ratio of the central peak to the ring peak) around the currently observed compact disk would have been detected, if they were present. 

A comparison between the GO Tau and V836 Tau disks provide an instructive example of the differences between a compact and large disk.  Both disks have inner emission cores with similar size and peak brightness.  Any 3-10$\sigma$ ring, like that seen in GO Tau, would have been easily detected in the outer disk of V836 Tau, if rings were present. If the disk brightness of GO Tau were scaled down by a factor of 2--3 to match the total disk flux of V836 Tau (as opposed to the peak flux), then the innermost bright ring is still detectable  when simulated with \textsc{CASA}, while the outer faint ring is barely visible. We cannot reject the possibility of very faint outer rings beyond our observational limit in some compact disks, perhaps especially the faintest disk, HQ Tau.

Our observations are only sensitive to substructures with scales of $\sim 10$ au. The non-detections of substructures in our compact disks (as well as the inner emission cores of ring disks) imply that any hidden substructure should be narrow or have low contrast. The three smallest disks ($\sim$30 au, DoAr 33, WSB 52 and SR 4) in the DSHARP sample \citep{huang2018_ring} have disk sizes that are similar to the radii of our compact disks. With a fine resolution of 5 au, radial profiles for DoAr 33 and WSB 52 show emission bumps instead of distinctive gaps, while SR 4 has a prominent deep gap around 11 au \citep{huang2018_ring}.
By convolving the DSHARP data with our beam size, the dust disks of DoAr 33 and WSB 52 become smooth, while the deep gap in SR 4 remains visible. 
In case of efficient dust trapping, dust rings are expected to have width equal to or narrower than the pressure scale height (e.g., 0.1, \citealt{dullemond2018}), thus substructures in the inner disk should have small characteristic scales.
The longest baselines of ALMA are needed to image the compact sources, probing the disk material distribution in the giant-planet forming region of our Solar System. 

\begin{figure}[!t]
\centering
    \includegraphics[width=0.45\textwidth]{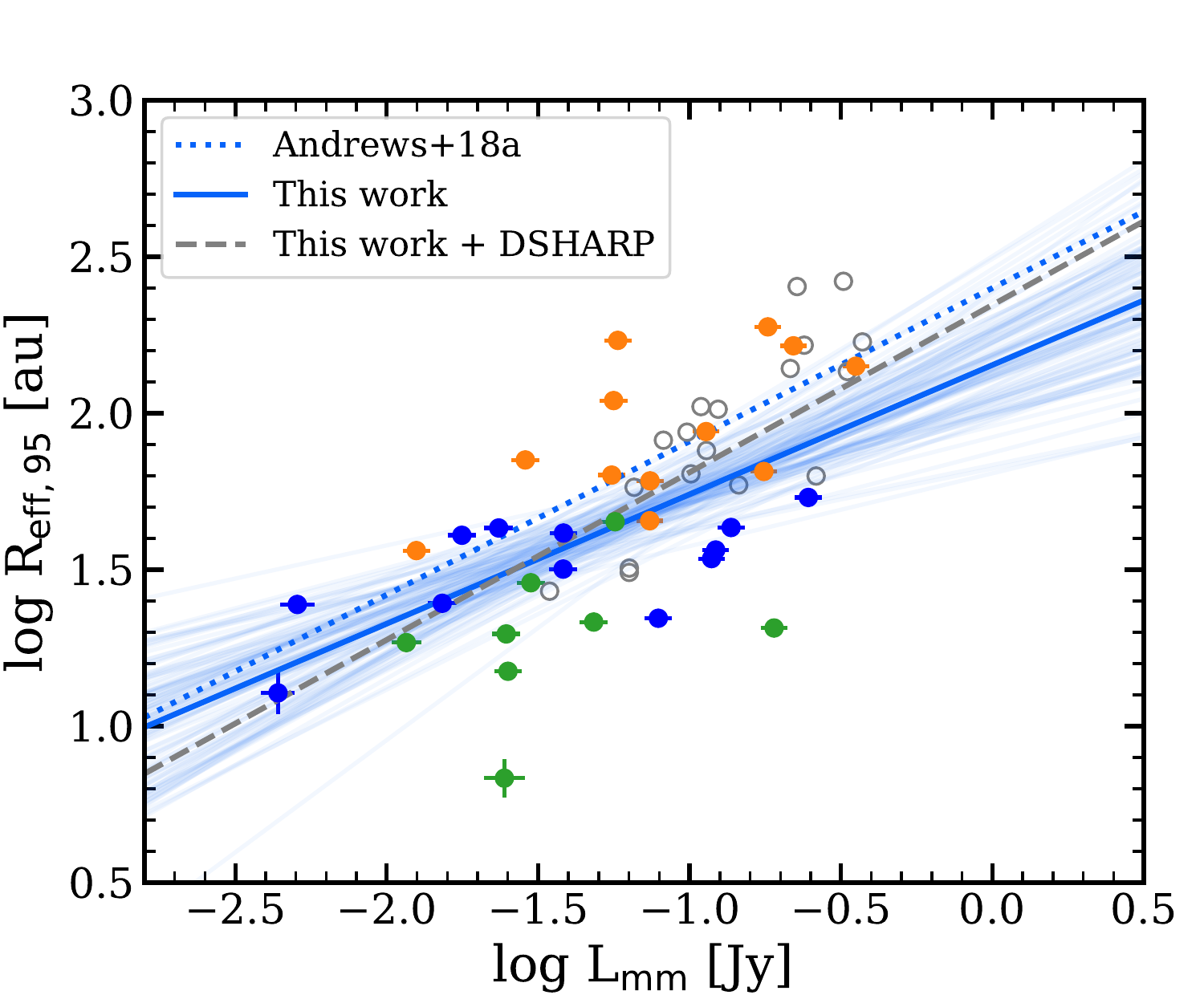} \\
    \caption{The disk continuum luminosity vs. disk radius. Colors are as in Figure~\ref{fig:ms_md_r} for different sub-samples and the DSHARP sample is shown in grey open circles. The solid blue line shows the linear regression analysis to our Taurus sample, with 100 random MCMC chains overlaid as light blue, while the grey dashed line shows the relation including the DSHARP sample. 
    \label{fig:Lmm_r}}
\end{figure}


\subsection{Disk size--luminosity relationship}
Disk population studies reveal scaling relations in multiple dimensions (e.g., $M_*, M_{disk}$, $\dot{M}_{*}$, $R_{disk}$), linking disk evolution with the bulk properties of disks (e.g., \citealt{manara2016}, \citealt{ansdell2016}, \citealt{pascucci2016}, \citealt{mulders2017}).  Recent analysis based on spatially resolved observations of 105 disks demonstrate that disk luminosity scales linearly with the surface area of the emitting materials \citep{andrews2018_Lmm}. With better mapping of the disk material distribution, we revisit this relationship to obtain a better understanding of disk demographics.

Figure~\ref{fig:Lmm_r} shows the resulting disk size--luminosity relation for our sample in the Taurus star-forming region. Assuming a linear relationship in the log--log plane, we employ the Bayesian linear regression method of \citet{kelly2007} with its python package \textit{Linmix}\footnote{\url{ https://github.com/jmeyers314/linmix}} to determine the correlation, considering uncertainties on both axes (including 10\% absolute flux uncertainty for luminosity). 
With this approach, we find a best-fit relation of log$R_{\rm eff}$ = (2.15$\pm$0.15) + (0.42$\pm$0.11)log$L_{\rm mm}$, where the disk size is the radius that encircles 95\% of flux, scaled the disk luminosity as $F_\nu(d/140)^2$ to a uniform 140 pc. The 1$\sigma$ dispersion is 0.3 dex and the correlation coefficient is $r = 0.58$. 
The slope of the relationship is consistent (1$\sigma$) with the finding in \citet{tripathi2017} and \citet{andrews2018_Lmm} that $L_{\rm mm} \propto R_{\rm eff}^2$. We also include the scaling relation from \citet{andrews2018_Lmm} with the same 95\% definition for disk size in Figure~\ref{fig:Lmm_r}, which shows a larger intercept (by 1-2$\sigma$). While our observations are conducted at 225 GHz, \citet{andrews2018_Lmm} used data from the $\sim$340 GHz band in their work. 
The lower intercept of our derived correlation might be caused by a more concentrated distribution of larger grains and finer angular resolution in our observations, or by the exclusion of some of large disks in our analysis. 
\citet{andrews2018_Lmm} also claims that the slope of the correlation is insensitive to the metrics (50\%--95\%) used for disk size definition, while we find a slightly flatter slope (0.34$\pm$0.11) when using $R_{\rm eff,68\%}$.

As seen in Figure~\ref{fig:Lmm_r}, disks with substructures mostly sit above the derived relationship. The inclusion of the DSHARP sample ($R_{\rm eff}$ defined as the radius of 95\% encircled), steepens the relation by 1$\sigma$ (0.53$\pm$0.08), while still keeping most of the ring disks in the top right of the plot.  \citet{tripathi2017} and \citet{andrews2018_Lmm} reproduce the scaling relation by considering optically thick disks with fractional regions being depleted (optically thin), and interpret the spread of the correlation as the varying fraction of the optically thin region. Taking a few disks with identical disk luminosity but different spatial extents, the most compact ones are likely to be optically thick overall, while the most extended ones are expected to contain large fraction of depleted optically thin regions (e.g., multiple gaps). This picture fits into the spatial segregation in the $L_{\rm mm}-R_{\rm eff}$ plane for disks with various morphologies. Multi-wavelength observations would be beneficial to access the spectral index information as to provide further evidence for this hypothesis.

\subsection{The origin of compact dust disks}
The continuum emission at millimeter wavelength is heavily dominated by dust grains at size of $\sim$ millimeter. 
These mm-size particles are subject to fast radial drift and are expected to be quickly depleted at large disk radii \citep{weidenschilling1977}; in contrast, dust disks often have large radii and survive for 1--10 Myr.  Disk substructures may resolve this apparent contradiction, serving as the mechanism (e.g., dust traps, \citealt{pinilla2012,dullemond2018}) to hinder radial drift and preserve the disk materials in wide orbits. Our observations seem to fit into this picture, where rings are formed at large radii and are the macroscopic consequence of  particle trapping, which helps to maintain a large population of mm-sized grains in the outer disk.
For our compact disks, the outer dust disk could be lost through efficient radial drift as dust rings are somehow not able to exist, as seen in the disk of CX Tau, which is very compact (and also smooth) in mm dust emission but has a very extended CO gas disk \citep{Facchini2019}.
In addiion, the compact disks may suffer from past dynamical interactions of very wide binaries, e.g. GI Tau and GK Tau with projected separation of 13.$\farcs$6 \citep{kraus2009wide}.

Alternatively, the compact disks could have small sizes initially, closely connected to the disk formation process. Non-ideal MHD simulations show that disk size distribution in early protostellar stage strongly depends on the relative orientation of the rotation vector of molecular cores and magnetic field \citep{tsukamoto2015,wurster2016}, through which both small and large disks could be formed.  However, the disk formation process remains unclear, with complications from initial angular momentum distribution, magnetohydrodynamic structure, and turbulence \citep{li2014,tsukamoto2018,Bate2018}.

The inclusion of disk gas size measurements is crucial for the assessment of the formation and evolution path of our compact disks. If the original outer regions of these compact disks are absent through rapid inward drift of mm-size grains, $R_{\rm gas}/R_{\rm dust, mm}$ should be higher in smaller disks. Disks that are born small would be expected with also small gas disks.

\section{Summary} \label{sec:summary}
We present a high-resolution ($\sim0\farcs12$) ALMA survey of 32 protoplanetary disks around solar-mass stars in the Taurus star-forming region. Our main goal is to provide an initial assessment of disk structures of mm-size grains at 10--20 au scale, for a sample of disks that spans a wide range in disk mm brightness. The disk model fitting is performed in the visibility plane to quantify the dust distribution. Our main results are summarized as follows: 

\begin{enumerate}

\item We detect disk substructures (including rings, gaps, and inner cavities) from 12 disks in the millimeter continuum emission. The other 20 smooth disks (without resolved substructures at current resolution, 12 disks in single stars, 8 disks around the primary star in multiple stellar systems with separations in $0\farcs7-3\farcs5$) are well described by an exponentially tapered power law profile, which may host unresolved small-scale substructures based on image residuals.  The non-detection of substructures in the smooth disks indicate that any hidden substructures are rather narrow or are low-contrast features.


\item Substructures are detected in disks around stars of all spectral types between A and M3, and in most bright and large disks. The subsample of disks with substructures has similar distributions as the smooth disks in stellar mass, stellar age, and disk luminosity.  However, the disks with subtructures have preferentially larger radii in mm-size grains. All disks with radius larger than 55 au show substructures in our sample.

\item The inner emission cores of the extended ring disks have comparable radius, peak brightness, power law index ($\gamma_1$), and taper index ($\gamma_2$) to the compact smooth disks. The large value of $\gamma_2$ in most of our disks may imply some level of radial drift of mm-size grains. The larger disk radii in the ring disks compared to the compact smooth disks is due to the presence of additional bright rings outside of the inner core.

\item The disk size--luminosity relation for our sample is broadly consistent with the correlations found by \citet{tripathi2017} and \citet{andrews2018_Lmm} from larger samples.  Some of the compact disks may be optically thick, while extended disks contain some regions that are optically thin, corresponding to the observed dust gaps in the large disks.

\item These compact smooth disks may have lost their outer disk through rapid inward migration, or they may still retain very faint outer disks that are below our sensitivity limit. Another possibility is that they were born small.  Future high-resolution observations toward low-mass stars and fainter disks will help to build a more complete picture of the occurrence and morphology of disk substructures, and facilitate a better understanding of the first steps toward planet formation.  

\end{enumerate}

\paragraph{Acknowledgments}
\acknowledgments{
We thank the Herschel/WISH team (PI: E. F. van Dishoeck), which hosted a team meeting at which this proposal idea was initially generated. We are grateful to Adam Kraus and Subo Dong for nice discussions.  F.L. and G.J.H. are supported by general grants 11773002 and 11473005 awarded by the National Science Foundation of China. 
P.P. acknowledges support by NASA through Hubble Fellowship grant HST-HF2-51380.001-A awarded by the Space Telescope Science Institute, which is operated by the Association of Universities for Research in Astronomy, Inc., for NASA, under contract NAS 5-26555. D.H. is supported by European Union A-ERC grant 291141 CHEMPLAN, NWO and by a KNAW professor prize awarded to E. van Dishoeck. C.F.M. acknowledges support through the ESO Fellowship and partial support by the Deutsche Forschungs-Gemeinschaft (DFG, German Research Foundation) - Ref no. FOR 2634/1 TE 1024/1-1. M.T. has been supported by the DISCSIM project, grant agreement 341137 funded by the European Research Council under ERC-2013-ADG and by the UK Science and Technology research Council (STFC). F.M., G.v.d.P and Y.B. acknowledge funding from ANR of France under contract number ANR-16-CE31-0013 (Planet-Forming-Disks). D.J. is supported by NRC Canada and by an NSERC Discovery Grant. BN and GL thank the support by the project PRIN-INAF 2016 The Cradle of Life - GENESIS-SKA (General Conditions in Early Planetary Systems for the rise of life with SKA). C.F.M., G.L., and M.T. have received funding from the European Union’s Horizon 2020 research and innovation programme under the Marie Sklodowska-Curie grant agreement No 823823 (DUSTBUSTERS).
Y.L. acknowledges supports by the Natural Science Foundation of Jiangsu Province of China (Grant No. BK20181513) and by the Natural Science Foundation of China (Grant No. 11503087). G.D. and E.R. acknowledges financial support from the European Research Council (ERC) under the European Union's Horizon 2020 research and innovation programme (grant agreement No 681601). N.H. thanks the LSSTC Data Science Fellowship Program, which is funded by LSSTC, NSF Cybertraining Grant 1829740, the Brinson Foundation, and the Moore Foundation; his participation in the program has benefited this work.

This paper makes use of the following ALMA data: 2016.1.01164.S. ALMA is a partnership of ESO (representing its member states), NSF (USA), and NINS (Japan), together with NRC (Canada), MOST and ASIAA (Taiwan), and KASI (Republic of Korea), in cooperation with the Republic of Chile. The Joint ALMA Observatory is operated by ESO, AUI/NRAO, and NAOJ. This work has made use of data from the European Space Agency (ESA) mission Gaia (https://www.cosmos.esa.int/gaia), processed by the Gaia Data Processing and Analysis Consortium (DPAC, https://www.cosmos.esa.int/web/gaia/dpac/consortium). Funding for the DPAC has been provided by national institutions, in particular the institutions participating in the Gaia Multilateral Agreement. 
This work used the Immersion Grating Infrared Spectrometer(IGRINS) that was developed under a collaboration between the University of Texas at Austin and the Korea Astronomy and Space Science Institute (KASI) with the financial support of the US National Science Foundation under grant AST-1229522, of the University of Texas at Austin, and of the Korean GMT Project of KASI.
These results made use of the Discovery Channel Telescope at Lowell Observatory. Lowell is a private, non-profit institution dedicated to astrophysical research and public appreciation of astronomy and operates the DCT in partnership with Boston University, the University of Maryland, the University of Toledo, Northern Arizona University and Yale University.
}

\software{Galario \citep{tazzari2018}, emcee \citep{Foreman-Mackey2013}, CASA (v4.7.2, v5.1.1; \citealt{McMullin2007})}

\appendix

\section{Disks not selected for this survey} \label{sec:source-selection}

In Section~\ref{sec:sample}, we briefly describe the source selection for this survey.  The initial source list was obtained from \citet{andrews2013}, which was compiled from the sample of disks identified in {\it Spitzer} imaging by \citet{Rebull10} and \citet{luhman2010}.  Using this source list, we first selected sources that are around stars with spectral types listed as earlier than M3 in \citet{herczeg2014}.  This restricted the sample to a reasonable size to answer our primary science questions.  Inclusion of later spectral types would have led to the selection of much fainter disks, which would not be feasible for a snapshot survey.

Binaries with separations of $0\farcs1$--$0\farcs5$ \citep{white2001} were excluded from our sample, because at those separations the binary interactions are expected to significantly affect disk substructures at the $\sim 0\farcs1$ resolution of our observations.  However, spectroscopic binaries were included in this sample because their spatial separations are not expected to affect large-scale disk structures.  

We excluded stars with extinctions $A_V>3$ mag and stars with faint $J$-band magnitudes.  In many of these cases, the objects have inner disks that are edge-on and block the light from the central star.   While disks viewed edge-on are powerful probes of disk flaring \citep{louvet2018}, substructures are challenging to identify and interpret (see, e.g., our observations of IQ Tau).  This criterion also selected against young disks, such as HL Tau, that are still embedded in remnant circumstellar envelopes, and other disks, such as FY Tau and IT Tau, that are highly extincted. IRAS 04216+2603 was excluded because of the high $A_V$ estimated by \citet{Rebull10}. CIDA 9 was included in our sample despite  faint 2MASS photometry because the source was bright and had low extinction in the \citet{herczeg2014} survey.  This selection criterion biases our survey against edge-on disks and perhaps favors disks that are slightly older.

Several disks, such as V819 Tau and JH 56, were excluded because their properties seem more similar to debris disks than primordial disks \citep[e.g.][]{hartmann2005,furlan2006}.  The mid-IR excess emission from these disks is very weak, and the stars show no sign of accretion.  These disks were not detected in the sub-mm by \citet{andrews2013}.  This exclusion means that we are not sensitive to what may be either the very last stages of disk evolution or the youngest debris disks.

From this list, we then selected targets that would not duplicate observations of disks that had been obtained or were scheduled for Cycle 3 observations\footnote{Based on the file duplication\_cycle4\_march18.xls located at 
https://almascience.eso.org/documents-and-tools/cycle4/duplication\_check\_xls/view.
} with a beam size of $<0\farcs25$.  This criterion ensured that our observations would be an improvement by at least a factor of 5 in the beam area.  However, 
this final selection criterion may introduce significant biases into our sample. Table~\ref{tab:excluded} lists all disks excluded because of duplication alone.  Several disks, including DM Tau, LkCa 15, and UX Tau, have known disk substructures inferred from mid-IR photometry, which in some cases have been previously imaged.  Some of these disks, including CW Tau, DG Tau, and DP Tau, drive prominent jets.

Our final selection then excluded a few sources for non-scientific reasons.
IRAS 04429+1550 and 2MASS J04333278+1800436 are located far from other sources and were excluded to maximize the efficiency of the program.  FP Tau was excluded based on a spectral type of M4 in \citet{luhman2010}, although \citet{herczeg2014} later re-classified it as M2.5.  Finally, the transition disk of GM Aur was included in our final sample in the proposal but was not observed, presumably because of inclusion in a different program.

Since this selection, additional disks have been found with {\it WISE} \citep{Rebull11,Esplin14}.  These new disks are typically located outside of the regions with highest stellar densities, since those had been covered by {\it Spitzer}.  The disks also tend to be around objects that are faint in near-IR photometry, either because the central object is a very low-mass star or brown dwarf or because the disk is viewed edge-on and obscures the central source.  However, some of these recently discovered disks would have likely been selected for this study, and their exclusion may introduce some bias in age or environment.

Some targets that have low disk masses for their stellar mass
are preferentially missing from our survey (see Figure~\ref{fig:ms_md_r}).  Our selection therefore appears biased.  However, many of these targets are close binaries.  For instance, V807 Tau is a $0\farcs3$ binary \citep{white2001} with a weak sub-mm flux from \citet{andrews2013}.  Disks in such multiple systems may sometimes have weak sub-mm fluxes  \citep[e.g.][]{harris2012,long2018-chaI} because of disk truncation or because the disk is around the fainter star.  Some weak sub-mm points, such as IRAS 04301+2608, are Class I objects \citep{furlan2011} and have only weak compact emission.  Some very high extinction objects, such as IRAS 04303+2240, may have edge-on disks.  If we discount the debris-like disks, then our selection samples well the full range of sub-mm flux from single and wide binary T Tauri stars with spectral types earlier than M3 in Taurus.

\section{Descriptions of disk host properties} \label{sec:source-detail}

Masses and ages of young stars are usually estimated by comparing their effective temperatures and luminosities to sets of evolutionary tracks.  The adopted temperatures and luminosities are discussed in \S \ref{sec:host-star}.  This approach is adopted here because it is easily reproduceable and applicable to the full set of Taurus objects.  Additional details for RY Tau and CIDA 9 are described in the subsections below.

The mass and age estimates are plagued by uncertainties, especially for K and M stars.  Our masses tend to be lower than the mass measured from Keplerian rotation of CO in the disk, consistent with direct comparisons for eclipsing binaries \citep{david2019}.  The adoption of the magnetic \citet{feiden2016} tracks would have produced higher masses, since the magnetic fields generate cooler atmospheres.   Spots are also not considered in either the observational measurements or in the evotionary tracks \citep[see, e.g.,][]{somers2015,gully2017}.


\subsection{RY Tau}
The stellar properties of RY Tau require a re-evaluation of the observed photospheric emission and the Gaia DR2 distance.  Initial spectral types of RY Tau of F8 and G0 were measured by \citet{Hubble22} and \citet{Joy45}, with independent support from \citet{Petrov99}, \citet{Calvet04}, and \citet{herczeg2014}, among others.  The spectral type of K1 measured by \citet{Herbig77} propagated into the \citet{Herbig88} catalog of T Tauri stars, and has since been widely adopted by many surveys and compilations \citep[e.g.][]{Kenyon95,Rebull10,luhman2010}.

To resolve this discrepancy in SpT, we downloaded (from the CADC archive) and coadded 96 high-resolution ESPaDOnS spectra \citep{Donati06} of RY Tau, obtained by PIs H. Takami, J.-F. Donati, and C. Dougados in separate programs, with some data publised by \citet{chou13}.  The spectra cover 3800--10000 \AA\ with a resolution of $R\sim68,000$.  We use the BT-Settl models \citep[version cifist2011\_2015;][]{Allard2014} with solar metallicity and $\log g=4.0$ to identify regions at $<4300$ \AA\ and 5150-5200 \AA\ that are most sensitive to temperature for FG spectral types.  A $\chi^2$ analysis on seven independent regions yields an effective temperature of $6220\pm80$ K, consistent with a spectral type F6-F8 in the \citet{Kenyon95} temperature scale.  The uncertainty of 80 K is the standard deviation of best-fit temperatures between several different spectral regions.  The discrepant spectral type of \citet{Herbig77} was likely caused by a measurement from a low-resolution red spectrum, which is not very sensitive to FG spectral types. Accounting for the minor change in spectral type from \citet{herczeg2014}, the extinction here is increased slightly to $A_V=1.94$ mag with an uncertainty estimate of $\sim 0.2$ mag.

The Gaia DR2 parallax leads to a distance of $445 \pm 45$ pc.  The 
uncertainty in parallax of 0.24 mas/yr is higher than the uncertainty of other nearby sources of similar magnitude, and the excess astrometric noise of 1 mas indicates a poor astrometric fit.   Bright nebulosity around RY Tau \citep[e.g.][]{Leavitt07} introduces additional uncertainty into whether the parallax measurement is reliable and likely causes the high excess noise.  For 29 Taurus members\footnote{Excluding the outlier parallax of IRAS 04158+2805, which is highly uncertain and may also be affected by nebulosity} listed in \citet{Esplin14} that are within 1 degree of RY Tau, the average distance is  $128.5\pm0.3$ pc, with a standard deviation of 5 pc.  If we focus on the 11 stars with the smallest uncertainties in parallax, the distance is $128.2$ pc with a standard deviation of 4 pc.  The distance of $128\pm4$ pc is adopted here as the distance to RY Tau.  

These updated parameters and the $J$-band brightness lead to a luminosity of 12.3 L$_\odot$ (the luminosity from \citet{herczeg2014} would be adjusted to 11.1 L$_\odot$).  Comparison to the non-magnetic \citet{feiden2016} yields a mass of 2.0 $M_\odot$, and an age of 5.2 Myr.


\subsection{CIDA 9}  
The 2MASS $J$-band is faint, relative to other Taurus sources of similar spectral type.  The $V$-band emission measured by the ASAS-SN survey \citep{kochanek2017} is highly variable, likely indicating extinction events.  The luminosity is therefore obtained directly from \citet{herczeg2014}.

The dynamical mass of CIDA 9 is $1.32\pm0.24$ M$_\odot$, as measured from CO rotation \citep{simon2017} and updated with Gaia DR2 distance.  This mass differs significantly from the mass of $0.43$ M$_\odot$ inferred from the spectral type of M1.8 \citep{herczeg2014}.  The inner hole of $\sim 25$ AU is suggestive of the higher mass.  However, the spectrum has strong TiO absorption and could not be mistaken for a K spectral type.  A sensitive search for spatially-resolved multiplicity revealed an absence of any close companion of CIDA 9A \citep{kraus2011}; the secondary is located at $2\farcs3$ and is discussed elsewhere.  

We recently obtained a high-resolution IGRINS \citep{Mace2018} HK spectrum of CIDA 9A to evaluate binarity.  The A component (in the SW) was placed on the slit.  The lines are single-peaked and located at a radial velocity of $\sim 18.7$ km s$^{-1}$, which corresponds to the expected velocity at that location in Taurus \citep{kraus2017}.  
For this one epoch (JD of 2458565.64), we can rule out that the source is a double-lined spectroscopic binary, although a robust test will require several epochs.  An initial analysis with a 2-temperature fit yields a photospheric temperature of 3800--4000 K, warmer than inferred from the TiO bands in the optical but cool enough to still be discrepant from the dynamical mass.

\begin{table}
\label{tab:excluded}
\caption{Disks excluded from this survey}
\begin{tabular}{ll}
\hline
Already observed: & AA Tau, AB Aur, CW Tau, CX Tau, CY Tau, DE Tau, DG Tau, DM Tau, DP Tau,  \\
& FX Tau, Haro 6-28, Haro 6-37, IS Tau, LkCa 15, SU Aur, UX Tau, V955 Tau, VY Tau \\
In our sample, not observed: & GM Aur\\
Excluded for $A_V>3$ mag: & V892~Tau, LR~1, JH~223, JH~112, IRAS~04260+2642, IT~Tau,
V410~X-ray~2, \\
~~~~or faint $J$-band: & IRAS~04301+2608,
IRAS~04370+2559,
IRAS~04385+2550,
FY~Tau,
CoKu~Tau~3, \\ 
& FZ~Tau,
IRAS~04303+2240,
IRAS 04125+2902,
V410 X-ray 7, IRAS 04196+2638,\\
 & 2MASS J04202144+2813491,
2MASS J04221675+2654570, GN Tau, \\
& 2MASS J04333905+2227207
 IRAS 04200+2759, MHO 3,
IRAS 04187+1927
V955 Tau, \\
& MHO~2,
MHO~1,
IRAS 04216+2603, ITG 33A, XEST 13-010, Haro 6-28
\\
Excluded for $0\farcs1-0\farcs5$ binarity: & V807 Tau, GG Tau, V955 Tau, CoKu Tau/4, IS Tau, GH Tau, FS Tau, \\
& IRAS 04187+1927, DF Tau, XZ Tau\\
Excluded for efficiency:&   IRAS 04429+1550, J04333278+1800436\\
Excluded due to use of prior spectral type: & FP Tau\\
\hline
\multicolumn{2}{l}{There might be some overlap in close multiples and high extinction targets}
\end{tabular}
\end{table}


\section{Fitting results for individual disks} \label{sec:note}
\begin{figure}[h]
\centering
    \includegraphics[width=0.95\textwidth]{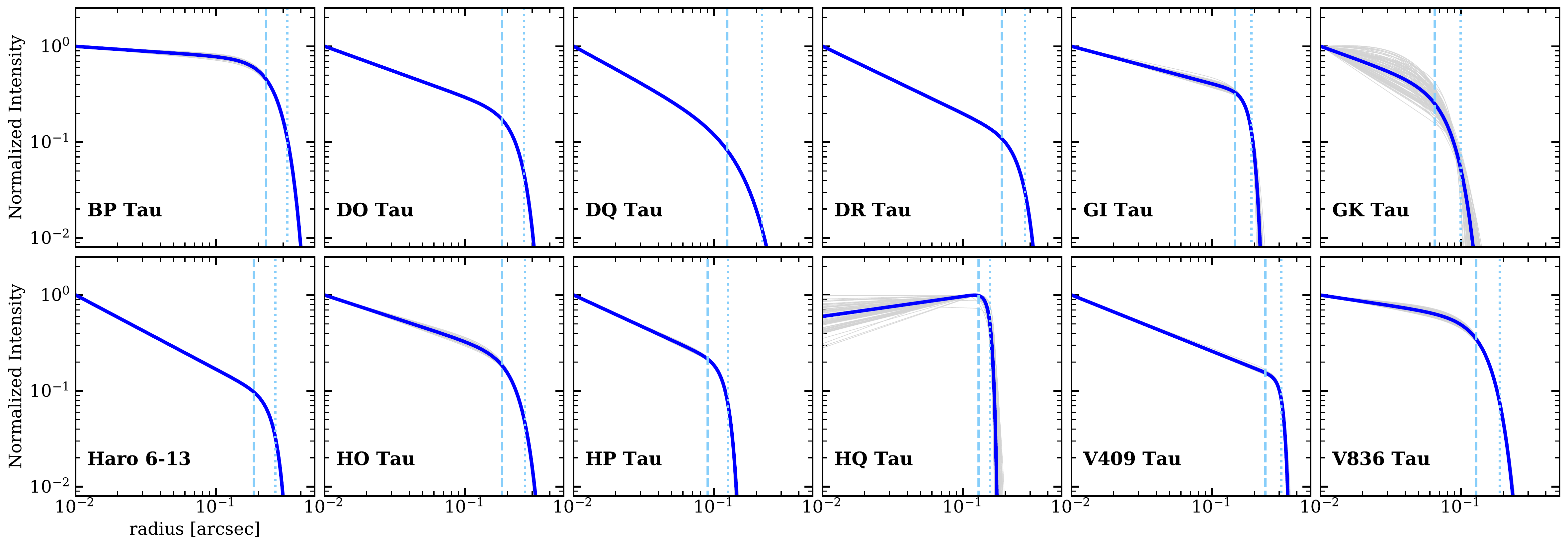} \\
    \caption{Best-fit intensity profile (red line) for the smooth single disks from the MCMC fits, with 100 randomly selected models from the fitting chains overlaid in grey. $R_{\rm eff, 68\%}$ and  $R_{\rm eff, 95\%}$ are labeled out in dashed and dotted line, respectively.   \label{fig:intensity_model}}
\end{figure}

\begin{figure}[h]
\centerline{\includegraphics[scale=0.9,trim=0 50 0 50]{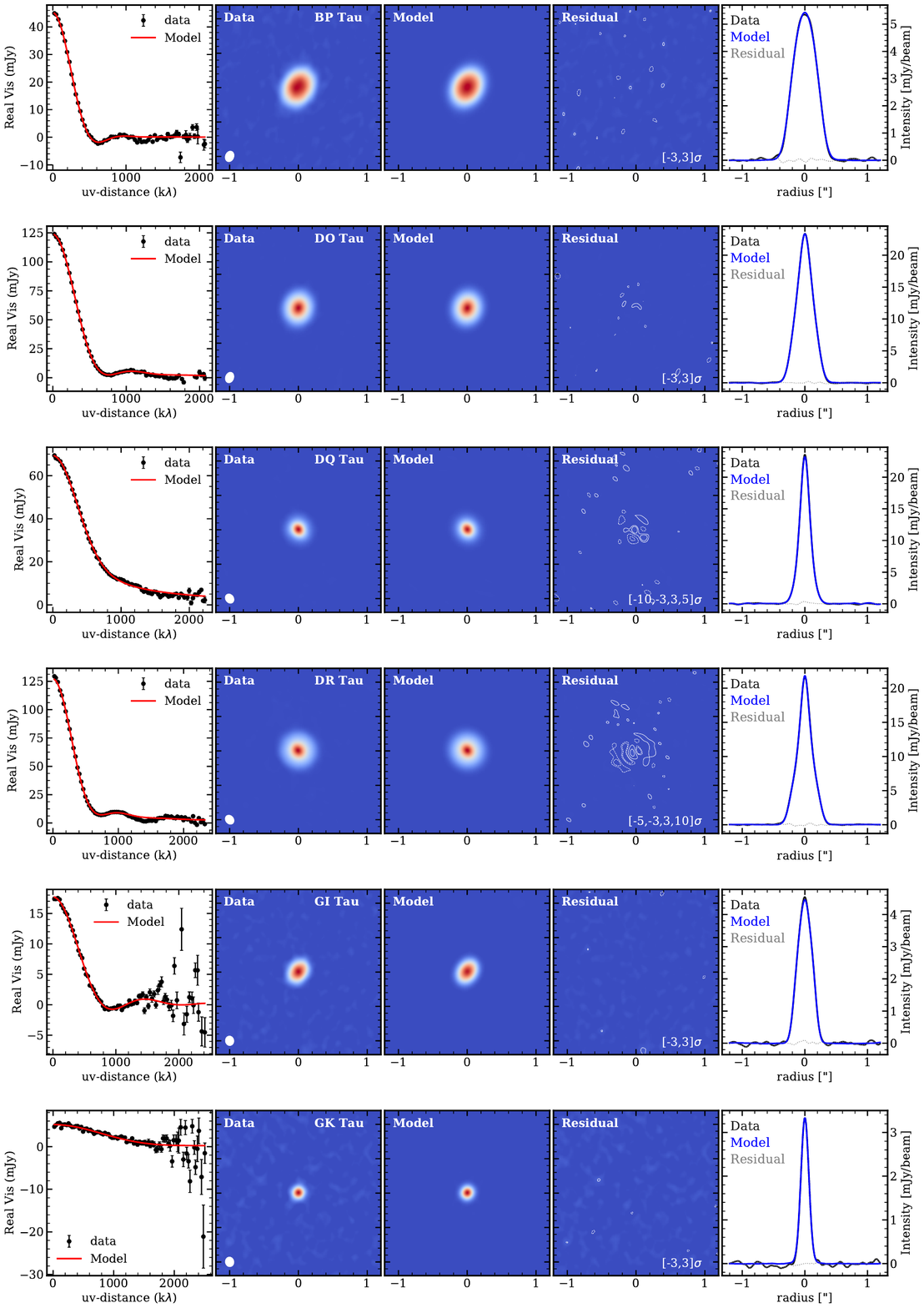}}
\captcont{A comparison of data and best-fit model for individual disk, including binned and deprojected visibility profile, continuum images (data, model, and residual maps), and radial profile along the disk major axis. \label{fig:model_result_single}}
\end{figure}

The best-fit model intensity profile for individual disks in the single smooth disk sample is shown in Figure~\ref{fig:intensity_model}.  
The comparisons of data and best-fit model are then shown in Figure~\ref{fig:model_result_single}, in which we check the goodness of our fit through visibility profile, synthesized image, and radial cut. In most cases, the maximum residual in the image is 3$\sigma$.

\clearpage
\pagebreak

\begin{figure}[h]
\centerline{\includegraphics[scale=0.9,trim=0 50 0 50]{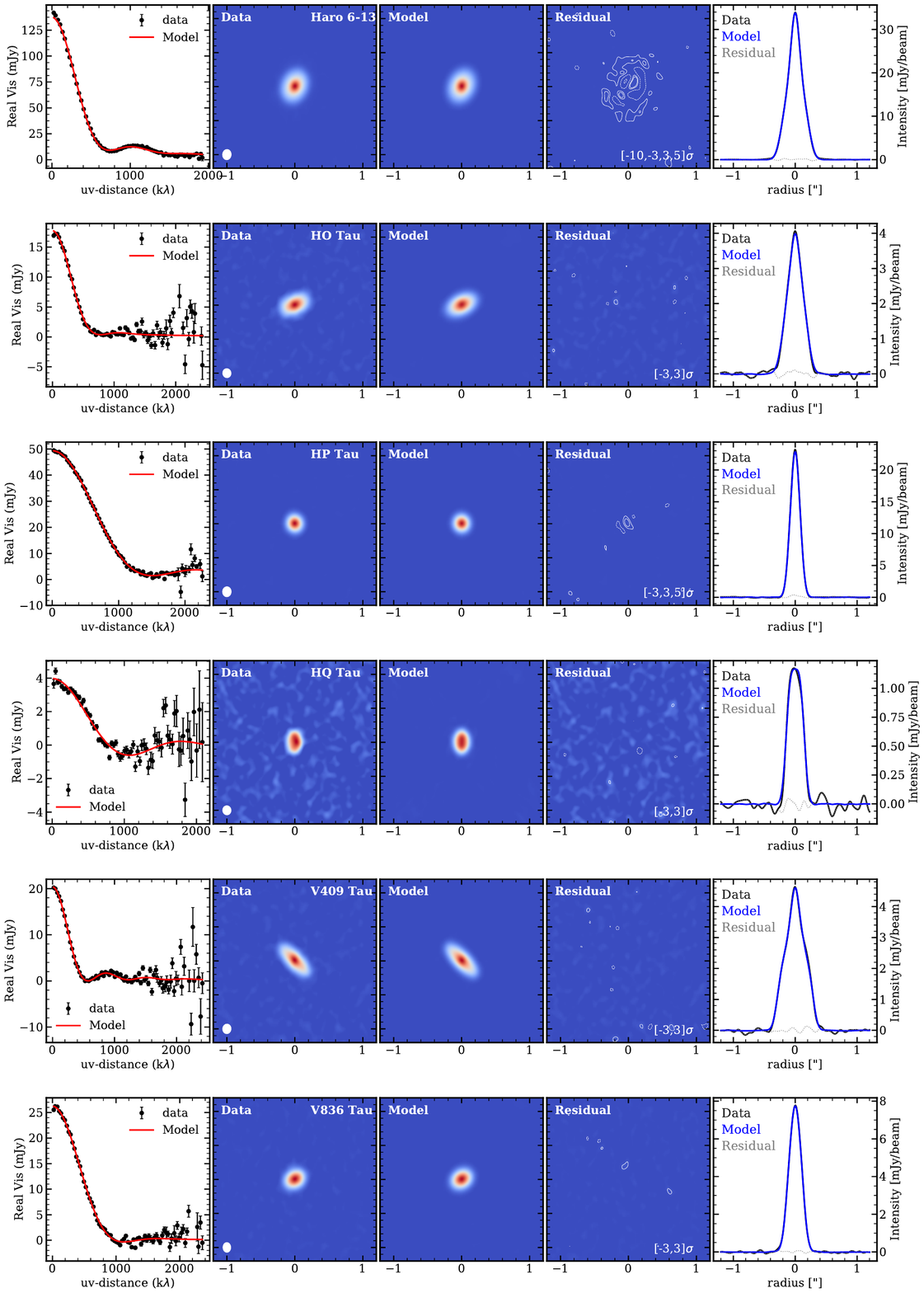}}
\captcont{Cont.}
\end{figure}

\begin{deluxetable*}{lcccccccc}
\tabletypesize{\scriptsize}
\tablecaption{Disk Model Parameters from Different Sets of Observations\label{tab:fitting_model_multiple}}
\tablewidth{0pt}
\tablehead{
\colhead{Name} & \colhead{$F_\nu$} & \colhead{$R_{\rm eff,68\%}$} & \colhead{$R_{\rm eff,95\%}$} & \colhead{$R_c$} & \colhead{$\gamma_{1}$} & \colhead{$\gamma_{2}$} & \colhead{incl} & \colhead{PA} \\
\colhead{} & \colhead{(mJy)} & \colhead{(arcsec)} & \colhead{(arcsec)} & \colhead{(arcsec)} & \colhead{} & \colhead{} &  \colhead{(deg)} & \colhead{(deg)}  
} 
\colnumbers
\startdata
BP Tau &  45.15$^{+0.19}_{-0.14}$ & 0.226 &  0.321 &  0.273 &  0.10$^{+0.03}_{-0.03}$  &  3.93$^{+0.24}_{-0.24}$ & 38.2$^{+0.5}_{-0.5}$  & 151.1$^{+1.0}_{-1.0}$\\
BP Tau SPW01  &  45.48 & 0.225 & 0.319 & 0.270 & 0.06 & 3.90 & 39.0 & 150.9 \\
BP Tau SPW23  &  44.31 & 0.223 & 0.317 & 0.270 & 0.10 & 3.92 & 38.1 & 150.7 \\
BP Tau SPW45  &  43.65 & 0.224 & 0.308 & 0.286 & 0.23 & 5.29 & 36.8 & 150.1 \\
\hline
GI Tau &  17.69$^{+0.25}_{-0.07}$ & 0.145 &  0.190 &  0.193 &  0.39$^{+0.05}_{-0.05}$  &  9.69$^{+5.56}_{-3.66}$ & 43.8$^{+1.1}_{-1.1}$  & 143.7$^{+1.9}_{-1.6}$\\
GI Tau SPW01  &  17.85 & 0.146 & 0.185 & 0.191 & 0.38 & 16.26 &  45.1 & 140.9 \\
GI Tau SPW23  &  16.83 & 0.145 & 0.186 & 0.192 & 0.42 & 15.07 &  43.1 & 142.1 \\
GI Tau SPW45  &  17.29 & 0.143 & 0.190 & 0.188 & 0.35 &  7.43 &  43.7 & 144.2 \\
\hline
HO Tau &  17.72$^{+0.20}_{-0.17}$ & 0.183 &  0.267 &  0.242 &  0.48$^{+0.05}_{-0.05}$  &  4.30$^{+0.76}_{-0.65}$ & 55.0$^{+0.8}_{-0.8}$  & 116.3$^{+1.0}_{-1.0}$\\
HO Tau SPW01  &  17.67 & 0.180 & 0.259 & 0.247 & 0.57 & 5.27 & 54.9 & 113.7 \\
HO Tau SPW23  &  17.21 & 0.176 & 0.253 & 0.232 & 0.42 & 4.59 & 54.3 & 116.8 \\
HO Tau SPW45  &  17.57 & 0.185 & 0.260 & 0.255 & 0.59 & 6.29 & 56.0 & 116.1 \\
\hline
V409 Tau &  20.22$^{+0.12}_{-0.18}$ & 0.239 &  0.311 &  0.324 &  0.59$^{+0.03}_{-0.03}$  & 16.11$^{+6.25}_{-5.98}$ & 69.3$^{+0.3}_{-0.3}$  &  44.8$^{+0.5}_{-0.5}$\\
V409 Tau SPW01  & 21.09 & 0.238 & 0.313 & 0.325 & 0.60 & 13.44 & 69.3 & 44.5 \\ 
V409 Tau SPW23  & 19.50 & 0.240 & 0.314 & 0.325 & 0.57 & 13.66 & 69.5 & 44.6 \\ 
V409 Tau SPW45  & 19.81 & 0.236 & 0.305 & 0.318 & 0.58 & 18.08 & 69.4 & 45.3 \\ 
\enddata
\tablecomments{For each disk listed here, modeling fittings were performed for three different sets of observations, and best-fit parameters from individual fits are listed as comparisons to values listed in Table~\ref{tab:fitting_model}.}
\end{deluxetable*}

\bibliographystyle{aasjournal}
\bibliography{ms}

\end{CJK*}
\end{document}